\documentclass[smallextended]{svjour3}  

\makeatletter
    \let\@internalcite\cite
    \def\cite{\def\citeauthoryear##1##2{##1, ##2}\@internalcite}
    \def\shortcite{\def\citeauthoryear##1{##2}\@internalcite}
    \def\@biblabel#1{\def\citeauthoryear##1##2{##1, ##2}[#1]\hfill}
\makeatother

\usepackage{listings}
\usepackage{array}

\usepackage{xspace}
\usepackage{multirow}
\usepackage{enumitem}
\usepackage{float}
\usepackage{color, colortbl}
\usepackage{booktabs} % For formal tables
\usepackage{pgfkeys}
\usepackage{graphicx}
\usepackage{caption}
\usepackage{subcaption}
\usepackage{booktabs,colortbl}
\usepackage{varwidth}
\usepackage{amsmath}
\usepackage{textcomp}
\usepackage{makecell}
\usepackage{siunitx, mhchem}
\usepackage{calc}
\usepackage{hyperref}

\newlength\myheight
\newlength\mydepth
\settototalheight\myheight{Xygp}
\newcommand{\udensdot}[1]{%
    \tikz[baseline=(todotted.base)]{
        \node[inner sep=1pt,outer sep=0pt] (todotted) {#1};
        \draw[densely dotted] (todotted.south west) -- (todotted.south east);
    }%
}%

\newcommand{\post}[1]{\href{{http://stackoverflow.com/questions/#1}}{\udensdot{#1}}}
\newcommand{\edit}[2]{\href{#1}{\udensdot{#2}}}
\newcommand{\footlink}[1]{\footnote{\href{#1}{#1}}}

\newcommand{\answer}[1]{\href{{http://stackoverflow.com/a/#1}}{\udensdot{#1}}}
\newcommand{\checkNum}[1]{#1}%{\color{orange}#1}}

\newcommand{\maintenance}{code maintenance }
\newcommand{\subscript}[2]{$#1 _ {#2}$}

\usepackage[absolute,overlay]{textpos}
\usepackage{tikz}
\usepackage{xcolor}
\usepackage{calc}

\renewcommand{\arraystretch}{1.2} 

\newcounter{findingctr}

\newcommand{\finding}[2]{\refstepcounter{findingctr}\emph{#1:\label{box:#2}}}

\newlength{\boxw}
\newlength{\boxh}
\newlength{\shadowsize}
\newlength{\boxroundness}
\newlength{\tmpa}
\newsavebox{\shadowblockbox}

\newenvironment{findingenv}[2]%
{\vspace{0.2cm}\noindent
\begin{lrbox}{
\shadowblockbox
}
\begin{minipage}{.98\columnwidth}
\finding{#1}{#2}~}%
{\end{minipage}\end{lrbox}%
\settowidth{\boxw}{\usebox{\shadowblockbox}}%
\settodepth{\tmpa}{\usebox{\shadowblockbox}}%
\settoheight{\boxh}{\usebox{\shadowblockbox}}%
\addtolength{\boxh}{\tmpa}%
\begin{tikzpicture}
\addtolength{\boxw}{\boxroundness * 2}
\addtolength{\boxh}{\boxroundness * 2}

\foreach \x in {0,.05,...,1}
{
\setlength{\tmpa}{\shadowsize * \real{\x}}
\fill[xshift=\shadowsize - 1pt,yshift=-\shadowsize + 
1pt,black,opacity=.04,rounded corners=\boxroundness] 
(\tmpa, \tmpa) rectangle +(\boxw - \tmpa - \tmpa, \boxh - \tmpa - 
\tmpa);
}

\filldraw[fill=white!50, draw=black!80, rounded corners=\boxroundness] (0, 
0) rectangle (\boxw, \boxh);
\draw node[xshift=\boxroundness,yshift=\boxroundness,inner sep=0pt,outer 
sep=0pt,anchor=south west] (0,0) {\usebox{\shadowblockbox}};
\end{tikzpicture}\vspace{0cm}%
}
\begin{document}
\title{On Using Stack Overflow Comment-Edit Pairs to Recommend Code Maintenance Changes}

\author{Henry Tang        \and
        Sarah Nadi
}


\institute{H. Tang \at
           University of Alberta \\
           \email{hktang@ualberta.ca}           
           \and
           S. Nadi \at
           University of Alberta \\
           \email{nadi@ualberta.ca}   
}

\date{Received: date / Accepted: date}

\maketitle

\begin{abstract}
%\todo{boast about accepted Prs in extremely popular repos}
%There are various types of code maintenance activities that developers perform.
%These include corrective maintenance, which typically involves bug fixing, as well as perfective maintenance, which includes improvements to the code.
%Various 
Code maintenance data sets typically consist of a before version of the code and an after version that contains the improvement or fix.
Such data sets are important for various software engineering support tools related to code maintenance, such as program repair, code recommender systems, or Application Programming Interface (API) misuse detection. 
%Many such data sets exist, but most focus on bug fixes that are mainly used for corrective maintenance activities.
%However, there are code changes that are important for other types of maintenance activities, such as perfective maintenance.
Most of the current data sets are typically constructed from mining commit history in version-control systems or issues in issue-tracking systems.

In this paper, we investigate whether Stack Overflow can be used as an additional source for building code maintenance data sets. 
%Stack Overflow has become an essential resource for software development.
Comments on Stack Overflow provide an effective way for developers to point out problems with existing answers, alternative solutions, or pitfalls.
Given its crowd-sourced nature, answers are then updated to incorporate these suggestions.
In this paper, we mine comment-edit pairs from Stack Overflow and investigate their potential usefulness for constructing the above data sets.
These comment-edit pairs have the added benefit of having concrete descriptions/explanations of why the change is needed as well as potentially having less tangled changes to deal with. 
We first design a technique to extract related comment-edit pairs and then qualitatively and quantitatively investigate the nature of these pairs.
We find that the majority of comment-edit pairs are not tangled, but find that only \checkNum{27\%} of the studied pairs are potentially useful for the above applications.
We categorize the types of mined pairs and find that the highest ratio of useful pairs come from those categorized as \textit{Correction}, \textit{Obsolete}, \textit{Flaw}, and \textit{Extension}.
These categories can provide data for both corrective and preventative maintenance activities.
To demonstrate the effectiveness of our extracted pairs, we submitted 15 pull requests to popular GitHub repositories, 10 of which have been accepted to widely used repositories such as Apache Beam\footlink{https://beam.apache.org/} and NLTK\footlink{https://www.nltk.org/}. 
Our work is the first to investigate Stack Overflow comment-edit pairs and opens the door for future work in this direction.
Based on our findings and observations, we provide concrete suggestions on how to potentially identify a larger set of useful comment-edit pairs, which can also be facilitated by our shared data.
\keywords{Stack Overflow, comment-edit pairs, bug-fix data sets}
\end{abstract}

%\begin{IEEEkeywords}
%TODO
%\end{IEEEkeywords}

\section{Introduction}
\label{sec:intro}

Software maintenance is an essential activity in the software development lifecycle.
In his seminal article, Swanson differentiated between three types of maintenance activities~\cite{swanson1976dimensions}.
\textit{Corrective maintenance} involves fixing faults in response to observed failures of the program (e.g., the shopping cart of a customer gets suddenly deleted during checkout).
\textit{Adaptive maintenance} involves changes needed to adapt the software to new data or processing environments, while \textit{preventative maintenance} is performed to improve the code such as eliminating processing
inefficiencies, enhancing performance, or improving maintainability.
All three types of \maintenance activities are necessary for projects to keep their code base up-to-date and ensure the system's quality on the long run. %e.g., through API deprecation from library updates.
We refer to any code changes that address the above maintenance categories as \textit{code maintenance changes}.

To support the above maintenance activities, many software engineering support tools such as defect prediction~\cite{fenton1999critique}, Application Programming Interface (API) misuse detection~\cite{amann2018systematic,amann2016mubench}, or program repair~\cite{GazzaloSWRepair}  were developed with the goal of automatically detecting, recommending, or applying code maintenance changes. 
To build or evaluate such tools, data sets of real code maintenance changes are needed.
The most common type of available data sets are \textit{bug-fix data sets} that are typically used for corrective maintenance, especially bug fixes (e.g.,~\cite{Just:2014:DDE,dallmeier2007extraction,cifuentes2009begbunch,amann2016mubench}).
%rely on bug-fix data sets, whether for building or evaluating the tools.
%A bug is defined as an error or fault in the program that causes unexpected and typically unwanted behaviour.\hk{maybe fine a reference for this definition}
While less common, there are also data sets that record code improvement changes related to perfective maintenance, such as using faster or more secure API calls~\cite{RaduNFBugs2019}.

All the above \textit{code-maintenance change data sets} (\textit{\maintenance data sets} for short) usually contain pairs of faulty/incorrect/low-quality code and the corresponding fixed/improved code, and are typically constructed from linking commits from version-control systems to reports in issue-tracking systems~\cite{Sliwerski2005}.
The commonly used linking approach relies on searching for commit messages that have specific keywords (e.g., fix, crash, hang, slow) and/or explicit links to issue IDs in issue-tracking systems~\cite{KimBugIntroChange,Nguyen:2012}.
While many widely used maintenance data sets have been constructed with this approach, relying on this linkage has several limitations: not all problems are documented in issue-tracking systems~\cite{Bachmann:2010}, not all developers are systematic about their linkage~\cite{Bird:2009,BissyandeBugLink}, and even worse, not every issue labeled as a bug is actually a bug~\cite{Herzig:2013Misclassification}.
Additionally, since the amount of code in a version control system is typically large and grouping separate code changes in a single commit (a.k.a \textit{tangled changes}~\cite{herzig2013impact}) is common~\cite{Herzig2016}, more advanced techniques that precisely identify the changes related to the maintenance activity of interest are required~\cite{Just:2014:DDE}.
%Traditionally, tangled changes are when a developer intends on performing a maintenance change (e.g., bug-fix), but includes unrelated changes that address other issues aside from the original issue.
Finally, finding good \textit{explanations} (i.e., a ``a reason or justification given for an action or belief'' based on the Oxford English dictionary) to attach to the identified maintenance activity, be it a bug fix or an improvement, such that they can be used in detection or recommender systems is difficult.
On one hand, commit messages are often short, meaningless, or non-descriptive~\cite{maalej2010can,dyer2013boa} and on the other hand, issue reports are often long with too many discussions~\cite{RastkarSummBugReport2010}.
Thus, the question is: are there complementary or additional sources of information that can be used to curate additional \maintenance data sets?
In this work, we investigate if Stack Overflow may be such a source. 

\label{intro:example}
Stack Overflow has become an essential resource for software developers.
It contains a wealth of information such as code solutions, best practices, and documentation of common pitfalls in response to the asked questions.
Given its crowd-sourced nature and high visibility as the go-to-place for information, Stack Overflow has the added advantage of community engagement where different developers point out various issues with the posted code snippets in the form of comments.
Comments may, for example, include pointing out faster APIs, missing version information, or simply wrong answers.
For example, Answer\footlink{http://stackoverflow.com/questions/50383046} 50383046 has a comment to include the \textit{rsplit} method as it is more efficient, a comment on Answer\footlink{http://stackoverflow.com/questions/19694159} 19694159 mentions the version differences of the answer between PHP pre 5.3 and after 5.3, and Answer\footlink{http://stackoverflow.com/questions/24261462} 24261462 has comments mentioning that the answer and even subsequent edit are incorrect.
The answerer, or other community members, then have a chance to edit the answer.
Stack Overflow records such changes in the answer edit history, including the code snippets contained in these answers.
Thus, if we can link comments to code-snippet edits, we can provide a new data source for building \maintenance data sets, which in turn can be used for the applications mentioned above, such as program repair or code improvement recommendations. 

Extracting comment-edit pairs from Stack Overflow can potentially address some of the problems discussed above: 
Stack Overflow code snippets are typically short and targeted, which overcomes the issue of tangled changes and removing unrelated code. 
Additionally, comments that result in an edit likely have the description of the issue that was addressed, which means that these comments can provide meaningful explanations that can accompany any code-change recommender tools.
%Inspired by the term \texiti{explainable artificial intelligence (XAI)} that aims to limit the black box nature surrounding the results of AI decisions by providing some explanation of the underlying AI decisions surrounding the results to humans, we also use the term \textit{explainable} recommendations or code changes.
%Instead of telling a developer 
%Similarly, comments from Stack Overflow can be used as an explanation for the paired code change recommendation.
For example, Answer\footlink{http://stackoverflow.com/questions/52517618} 52517618 contains code that converts a byte array to a string as follows \texttt{String s = new String(bytes, ``UTF-8'');}. This code snippet then gets updated to 
\texttt{String s = new String (bytes, StandardCharsets.UTF\_8);}.
If a code improvement/recommendation tool suggests this change to a developer, the developer may be unsure as to why this change is necessary and may end up ignoring the suggestion.
If, however, the following comment \textit{``On Java 7 you can also use \texttt{new String(bytes, StandardCharsets.UTF\_8);} which avoids having to catch the \texttt{UnsupportedEncodingException}''} is provided to the developer along with this change, they will understand the reasoning behind the suggestion and make an informative decision on accepting the suggestion. 
In short, a tool that detects the former (pre-edit) piece of code could suggest the latter (post-edit) piece of code and accompany that suggestion with the related comment to explain why the suggestion was made.

To explore the feasibility of using Stack Overflow as a source for \maintenance data, we first need to design a technique that maps comments to their corresponding edits. In other words, we need to extract \textit{comment-edit} pairs, i.e., a comment and the resulting edit that addressed this comment.
%Currently, there is no way to automatically identify why an edit to an answer has occurred.
%While an editor can provide a message to describe the edit they performed, most editors do not supply this edit message.
%We believe that identifying \textit{comment-edit} pairs, i.e., a comment and the resulting edit that addressed this comment, is valuable for several applications:
%(1) IDE-integrated API misuse/improvement detection systems that warn users about mistakes or provide improved alternative code snippets, (2) a Stack Overflow answer-edit detector system where answer posters can be directly warned about possible flaws in their code without waiting for comments, (3) a Stack Overflow tagging/labeling system that can explicitly mark suspicious or wrong answers such that end users are aware of the problems and active community users can be motivated to fix it. The last point aligns well with Stack Overflow's gamification approach. Similar to how bounties are placed for answering certain questions, a bounty (or other motivation/gamification mechanism) can be placed for updating wrong or outdated answers that have exceeded a certain time limit.
To do so, we leverage the SOTorrent~\cite{BaltesSOTorrent} data set and adapt and improve a previous matching approach we designed to identify ignored comments~\cite{SoniMSR19}.
At a high level, our  automated approach matches a comment to an edit if the comment occurred before the edit, the comment mentions a code term that gets added to or removed from a code snippet in the edit, and the commenter and editor are different users.
To support our investigation of using these comment-edit pairs for creating \maintenance data sets, this paper then answers the following research questions:

\begin{enumerate}[label=\subscript{RQ}{\arabic*},leftmargin=*]
\item \label{rq:precision} \textit{What is the precision of an automated technique for extracting comment-edit pairs from Stack Overflow?} There is currently no way on Stack Overflow to relate a comment to an edit, so the first step of this research is to establish an automated technique for doing this pairing, and to evaluate its precision.
\item \label{rq:tangled} \textit{How tangled are the changes in Stack Overflow  comment-edit pairs?} To investigate if the identified comment-edit pairs do indeed overcome the challenge of tangled changes, we investigate how often do the changes in mined pairs address issues other than that pointed out in the related comment.
\item \label{rq:changetype} \textit{What type of changes occur in Stack Overflow comment-edit pairs?} To understand what potential types of data sets and related software engineering applications can these comment-edit pairs be used for, we need to understand the types of changes that occur in them (e.g., syntax error fixes vs. catering the solution to the original poster's question).
\item \label{rq:usefulness}\textit{What is the potential usefulness of the extracted comment-edit pairs for curating \maintenance data sets?} Not all the mined comment-edit pairs are necessarily useful for \maintenance data sets. Thus, it is important to understand how many of the comment-edit pairs are useful for the intended applications. We consider a comment-edit pair as \textit{useful} for code recommender systems\footnote{Note that we use the term \textit{code recommender system} as a general umbrella for any support tool that suggests fixes, code changes, or related code snippets.} if (1) the edit addressing the comment happens to an existing code snippet in the answer such that there is code to be matched in a target system and (2) if the comment describes this change in a way that is understandable in isolation of the posted Stack Overflow question. We also investigate how tangled these useful pairs are, and which categories they fall under. To further demonstrate usefulness, we also submit \checkNum{15} pull requests based on our mined pairs to \checkNum{15} different open-source repositories.
\end{enumerate}

To answer the above research questions, we run our automated matching technique on five popular Stack Overflow tags (Java, JavaScript, Android, PHP, and Python). 
We then manually analyze a statistically representative sample of \checkNum{1,910} detected comment-edit pairs to confirm true matches.
We record the type of suggestion and change(s) being made, the presence of tangled changes in the edit, and the usefulness of the pair for the \checkNum{1,482} confirmed pairs we find.

Our results show that the precision of our automated approach is \checkNum{74\%-80\%} across the five tags and that only \checkNum{11\%} of the \checkNum{1,482} confirmed pairs are tangled, while \checkNum{27\%} are useful.
To categorize the confirmed pairs, we use a coding guideline from previous work~\cite{ZhangSOComments2019} that analyzed the types of comments on Stack Overflow but did not looking at corresponding edits.
We find that \checkNum{34\%, 16\%, and 13\%} of the confirmed pairs are of types \textit{Error}, \textit{Request}, and \textit{Correction} respectively, collectively consisting over 50\% of the confirmed pairs.
However, when looking specifically at useful pairs, we find that types \textit{Correction}, \textit{Obsolete}, \textit{Flaw}, and \textit{Extension} are the most useful.
This is promising for future maintenance applications as these types of comments are relatively more general and the corresponding edits will be applicable in a general setting.
Additionally, \checkNum{10} out of the \checkNum{15} pull requests we submitted based on our collected data have already been accepted.
These repos include popular and influential projects, such as Apache Beam\footlink{https://beam.apache.org/} and NLTK\footlink{https://www.nltk.org/}, which demonstrates the potential impact of our comment-edit pairs. 
%\sn{any updates on the remaining PRs? have these numbers changed?} \hk{10 accepted, 3 rejected, 2 probably won't be responded to}
To the best of our knowledge, this is the first work that maps Stack Overflow comments to edits and studies the potential of using these comment-edit pairs for constructing \maintenance data sets that also provide explanations for the provided changes. The summary of our contributions in this paper are as follows. 

\begin{itemize}
\item We implement an automated approach for matching comments to edits. We apply the approach to Stack Overflow posts covering five popular tags (Java, JavaScript, Android, Python, and Php) and extract a total of \checkNum{248,399} comment-edit pairs.
\item We manually analyze \checkNum{799} comments from 100 answers (20 from each of the five tags) to create a ground truth of \checkNum{194} comment-edit pairs, and use it to evaluate our matching approach and compare it to a naive baseline.
\item We manually analyze a statistically representative random sample of \checkNum{1,910} comment-edit pairs and confirm true matches for \checkNum{1,482} pairs. We record the category the comment belongs to, the presence of tangled changes, as well its usefulness for \maintenance data sets.
\item Based on the above collected data, we answer four research questions to determine if comment-edit pairs can be used in future maintenance-related software engineering applications. We also discuss challenges and opportunities for future work in this direction.
\item For additional external validation, we use the confirmed comment-edit pairs to submit \checkNum{15} pull requests to different open-source GitHub repositories. To date, \checkNum{10} of these pull requests have been accepted.
 \end{itemize}

All our code and data are publicly shared on our artifact page~\cite{artifact}.

\section{Related Work}
\label{sec:related}

We discuss two categories of related work.
The first is existing code maintenance data sets and the second is previous work that leverages data from Stack Overflow.

\subsection{Existing Code Maintenance Data Sets}
\label{sec:existing-dataset}

Over the last two decades, there has been a tremendous effort and movement towards curating useful data sets that can assist in maintenance tasks~\cite{menzies2012promise}, especially those related to corrective maintenance. We discuss a subset of the most relevant ones here.

iBugs~\cite{dallmeier2007extraction} was early work that uses the technique of identifying bug-fixing commits through keywords in commit messages. It collected pairs of before (buggy) and after versions (fixed) of the code along with the associated test suite. 
Defects4J~\cite{Just:2014:DDE} is a well-known data set of Java bugs that was built by mining version-control systems containing commit messages that explicitly reference a bug ID in the issue tracking system, or if a bug issue references a commit in the version-control system.
The data set contains two versions of the code, one before and one after the fix.
%One special feature of Defects4J is that in addition to this code pair, it also stores a reduced test suite that exposes the bug. 
Different from iBugs, Defects4J does some filtering of the test suite to keep only tests that fail on the buggy version and pass on the fixed version. 
To overcome the problem of tangled changes~\cite{herzig2013impact}, the authors manually reviewed the source code diffs of the before and after versions of the code and, if necessary, removed any irrelevant changes.

%Herzig et al.~\cite{Herzig:2013Misclassification} manually verified a set of 7,401 tickets from the issue-tracking systems of five projects. 
%The authors found that only 2,914 of these tickets actually report bugs.
%While the main focus of that work was to quantify the impact of issue misclassifications (i.e., a bug ticket not actually being about a bug) on bug prediction, the authors do publicly share their data set.
%However, this data set contains only the issues and their classifications and further processing is needed.

Dit et al.~\cite{dit2013dataset} again mined change history, linking commits to issue IDs to curate a data set that can be useful for software maintenance tasks. 
Their goal was for this data set to be useful for various maintenance tasks such as feature location, impact analysis, developer recommendations, and traceability recovery; however, they did not provide a categorization of the entries in their data set, so we are not aware of the exact maintenance tasks supported and their distribution.
Additionally, while both  our work and theirs target software maintenance, our extracted data is focused on \textit{code} maintenance activities, rather than more general tasks such as developer recommendation.

Ohira et al.~\cite{ohira2015dataset} manually categorized issue reports to identify high-impact bugs.
While they considered issues labeled as both BUG and IMPROVEMENT, they mentioned that most of the improvements are actually considered as bugs.
None the less, we assume that their data set may also be applicable to perfective maintenance activities, and not only corrective maintenance.
%To evaluate a proposed automated bug-fixing approach, Gao et al.~\cite{gao2015fixing} curated a set of 24 Android crash bugs by again mining reports from issue-tracking systems and manually analyzing them. 
The recent BugHunter data set~\cite{FERENC2020110691} again relied on issue trackers and commit history.
Different from other data sets, it tried to reduce the code changes in the before/after versions of the code in order to identify the minimal set of affected code elements.

While following similar methods of relying on commit messages and manually reviewing the changes, Radu and Nadi~\cite{RaduNFBugs2019} specifically focused on non-functional bugs that are related to aspects such security, performance, memory management, etc. 

BugSwarm~\cite{tomassi2019bugswarm} is a recent effort that attempts to remove some of the manual effort involved in curating bug-fix data sets. 
While it also relies on version-control history, it leverages the continuous integration (CI) service in the target repositories to identify bug-fixing commits through their CI build status.
Additionally, BugSwarm containerizes the before and after versions of the code and build scripts to ensure fully reproducible problems.

\paragraph{Summary} To summarize, most existing code maintenance data sets seem to focus on corrective maintenance tasks, specifically bug fixes. 
Additionally, most of these data sets are constructed by mining version-control history or issue-tracking systems. 
As mentioned in the introduction, this construction technique has been criticized because of missing problems in issue-tracking sytems~\cite{Bachmann:2010}, lack of systematic linking between commits and bug reports~\cite{Bird:2009,BissyandeBugLink}, misclassification in issue-tracking systems~\cite{Herzig:2013Misclassification}, and tangled changes not related to the fix~\cite{herzig2013impact,Herzig2016}. 
Our work is an attempt to find another data source for code maintenance data sets other than version-control or issue-tracking systems.
Additionally, since we do not limit ourselves to keywords such as ``fix'' or links to bug issues, using Stack Overflow may potentially provide changes related to additional code maintenance activities.
In general, our goal is not to replace or compete with current data sets, but instead to explore the potential of using Stack Overflow for curating additional relevant data sets.

\subsection{Stack Overflow Studies}
\label{sec:so-studies}

Data from Stack Overflow has been used extensively in previous work with varying purposes.
While some papers focus specifically on studying various characteristics of Stack Overflow and how information evolves on it~\cite{EmpricalStudyOfObsoleteAnswers,ZhangAdaptingCodeExamples,Barua2014}, others use information from Stack Overflow for specific purposes such as augmenting documentation, code search, or improving code analysis tools~\cite{rahman2019automatic,TruedeICSE16,Subramanian:2014,Ponzanelli:2014,Lin:2019:PMO,LiuSANER18}.
Given the nature of our work, which establishes a relationship between comments and code edits on Stack Overflow and investigates the nature of these pairs, in this section, we focus only on related work that studied/used comments or edits on Stack Overflow (SO).

\paragraph{Related work we rely on.}
Our previous MSR challenge paper~\cite{SoniMSR19} quantified how often comments cause answer updates, and how often comments are ignored even when they should have resulted in an answer update.
We used three heuristics for matching comments to edits and categorizing them: (1) code checks where a comment caused an update if a code element in the comment is added or removed in the edit, (2) keyword phrase checks that suggest that the comment is explicitly asking for an edit but no edit occurred, and (3) question checks where a comment explicitly asks a question about the posted code. 
Our results showed that code checks resulted in the most matches between comments and edits and that most of the wrongly labeled pairs occurred when we tried to deduce that a comment should warrant an update and was ignored, or that a comment does not warrant an update.
Based on these findings, in this paper, we only use the code check heuristic and focus on finding comment-edit pairs where an update actually occurred.
This current paper differs from our previous work in terms of goals: we do not try to automatically categorize \textit{all} comments and do not look for ignored comments.
Our goal is to find comments that actually caused an edit, and to study the comment-edit pairs in terms of their suitability for creating \maintenance data sets.
%We note that they studied only a subset of the answers in the 5 tags they focused on, while we run our analysis on all answers in those tags.
Additionally, we improve the matching algorithm and evaluate it against a manually constructed ground truth. We also manually validate a statistically representative sample of the pairs our tooling detects, measure the precision, and publicly share a validated data set containing the confirmed pairs.

Another recent work we rely on is that by Zhang et al.~\cite{ZhangSOComments2019}.
In that work, the authors analyzed comments on Stack Overflow. 
They investigated the information discussed in comments and performed open coding to categorize the analyzed comments.
They defined seven broad categories and 17 sub-categories of comments.
They did not, however, attempt to match comments to edits or analyze the code changes in edits.
Given that the comments we find in comment-edit pairs are a subset of all comments on Stack Overflow, we use the categories they create as our coding guideline for categorizing comments in our pairs. 
In other words, given Zhang et al.'s categories, we perform closed-coding (i.e., when codes/labels are predetermined) to categorize our comment-edit pairs.
Some of the categories of comments they found, such as pointing out errors or weaknesses in answers or providing alternative solutions, give us assurance that finding the edits corresponding to these comments can potentially be useful for \maintenance data sets.

%We categorize comments on the reasons an edit was made, only looking at comments that have already been marked as causing an update, not categorizing comments based on whether it did or should have caused an update.
%Soni et al.~\cite{SoniMSR19} also groups potential comments and edits based on their creation date, only comments with a creation date before an edit but after the previous edit are considered.
%However, this paper makes no such groupings and looks at all edits after a comment.
%e.g., with $c$ denoting a comment and $e$ denoting an edit, given an event ordering of \textit{init}, $c_1$, $e_1$, $c_2$, $c_3$, $e_2$, $e_3$ Soni et al.~\cite{SoniMSR19} would only consider $c_2$ and $c_3$ for $e_2$ and $e_3$ would have no comments. This paper would consider $c_1$, $c_2$, and $c_3$ for both $e_2$ and $e_3$, note that we do not allow a comment to be matched with muliple edits.
%Soni et al.~\cite{SoniMSR19} also relied on three heuristics to determine if a comment caused an edit: code checks, keyword checks, and question checks.
%Code checks looked for matching code in the comment and edit.
%Keyword checks used regular expressions to find comments that had no explicit code but matched a particular comment pattern. e.g., "This answer needs to be updated"
%Question checks also used regular expressions to find the pattern of "Interrogative word -- verb -- code" to mark the comment as "should have cause an update".
%In contrast, this project uses only code checks.
%If there is similar code in both the comment and edit then we deem it as having caused the edit. 

\paragraph{SO for Error Fixing.} Wong et al.~\cite{wong2019pythonsyntax} studied edits to Python code snippets on Stack Overflow in order to produce a syntax error data set. 
Their goal was to make a free, open, and public data set that would be representative of the kinds of syntax errors general developers would have.
At a high level, they parse the before and after versions of the most recent edit in an answer.
%Their methodology includes creating an abstract syntax tree (AST) on edits of the answer.
%As there are answers that do not contain valid Python code, Wong et al.~\cite{wong2019pythonsyntax} attempted to parse an AST on both the most recent edit and the one prior.
If the prior version included a parse error and the most recent did not, then they store the two versions as a syntax error and fix respectively. 
Our work differs as we focus on linking comments and edits to attach a reason for an edit.
We also do not focus solely on syntax errors and find changes related to more code maintenance activities, including various types of fixes and code improvements.

Thiselton et al.~\cite{thiselton2019compiler} used Stack Overflow answers in order to provide better compiler error messages for active development.
Their work takes a Python compiler error message and constructs a Stack Overflow query. 
They take the first question on the first page that is returned by the query that contains at least one answer.
They then take the accepted answer (or highest voted answer if there is no accepted answer) and modify the compiler error to incorporate a summary of the answer they found.
They do not use comments or edits on a Stack Overflow answer at all. 
However, their work highlights that novel applications using information from Stack Overflow can be useful in helping developers during active development.

Gao et al. proposed an automated bug-fixing approach that relies on mining information from Stack Overflow~\cite{gao2015fixing}, but they rely neither on answer edits or comments.
Instead, they find answers that contain two code snippets and rely on heuristics to identify the buggy and correct version (e.g., Instead of \texttt{code snippet X}, use \texttt{code snippet Y}). 
Alternatively, they try to match the buggy code snippet in the question to a modified, and presumably correct, code snippet in the answer. 
After matching these two versions, they use GumTree~\cite{DBLP:conf/kbse/FalleriMBMM14} to generate edit scripts for automated bug fixing.
While our sources of data are different, we foresee that future work can apply their automated edit  script generation technique to the pre/post pairs we extract.

\paragraph{Collaboration Characteristics on SO.} Adaji et al.~\cite{AdajiEditComment} also studied edits and comments on Stack Overflow.
Unlike our work that analyzes the contents of comments and edits to link them together, their work used comments and edits to study collaboration characteristics on Stack Overflow with the goal of finding the types of users that contribute to high quality answers.
Specifically, they investigated whether the number of comments on an answer or the reputation of the editor are correlated with the answer quality.
%The reputation of an editor being the number of answer related badges they have.
Their results showed that most of the edits made were by users with no badges and that most high quality answers had more comments rather than less.
Based on these findings, we study all comments and edits, regardless of the reputation of the user or the score of the answer.

Wang et al.~\cite{EditBadgeGamification} studied Stack Overflow badges that are related to revisions of answers.
They found that most revisions were made in spikes (i.e., many revisions made on the same day) rather than spread out over different days.
These spikes coincided with the days Stack Overflow were awarding badges to members, and the corresponding revisions during these spikes were mostly simple revisions (i.e., typo correction and formatting).
They also noted that most of the revisions made on these days needed to be rolled back due to the revision being incorrect or undesired.
They concluded that the current system of using badges was insufficient in enforcing answer quality and that there needed to be a change in how Stack Overflow encourages revisions without lowering the quality of answers.
Our work focuses on the contents of the revisions and relating them to comments, as opposed to motivation schemes for performing the edits.

\paragraph{Answer Quality.}\label{sec:answer-quality} Dalip et al.~\cite{Dalip:2013:EUF:2484028.2484072} created a learning to rank approach with the goal of automatically estimating the feedback a user would give regarding the quality of an answer.
To do so, they extracted features related to both comments and edits.
All their features are quantitative (e.g., number of edits, number of comments, or number of users who commented on answer), and they did not analyze the content of the comments or map comments to edits.
%However, they do not focus on finding the comment that caused the edit; instead they were interested in how an answer was improved.
%Dalip et al.~\cite{Dalip:2013:EUF:2484028.2484072} focused on how answers were enhanced. 
%e.g., edits to correct mistakes, include examples, or link to additional information, etc. 
%Based on that, they developed a better method to rank the answers compared to the current manual upvote system.
%They did not focus on finding the reasons or causes to those enhancing edits, but rather focused on the contents of the edit and how the answer was refined.

Diamantopoulos et al.~\cite{Diamantopoulos:2019:TMA:3341883.3341922} analyzed answer edits to determine what makes an optimal answer.
With that information, they discuss future Stack Overflow tools that could suggest edits on an answer to improve its quality.
While our work can help with similar future goals, the methodology and the focus of both studies differ substantially.
Diamantopoulos et al.~\cite{Diamantopoulos:2019:TMA:3341883.3341922} used a neural network to study the edits made on Java answers and applied clustering to extract related edits. 
They then used the ``commit'' message associated with an edit\footnote{note that they refer to this message as \textit{comment} in their paper, but it is not a comment on the answer, but rather the message the editor provides with their edit} to come up with representative descriptions for each cluster; however, as they also point out, having a message associated with the edit is rare.
Since comments on an answer are much more common and are also more descriptive, we believe that studying answer comments to understand the types of edits that occur may provide more explanations and intuitions for answer edits, which would make any follow up recommender system more useful to users. 
Additionally, we pair comments with the corresponding edits while they do not.
%It is worth noting that while we study more programming languages than Diamantopoulos et al., we limit our work to code edits to enable more precise matching while they look at all edits, including edits to text blocks.

Ragkhitwetsagul et al.~\cite{ToxicCode2019} studied the quality of Stack Overflow answers and found that many answers were outdated, buggy, incorrect, etc.
They also raise the issue that many answers also violate licensing as most answers are copy-pasted from users' own work.
While general Stack Overflow answer quality is a concern, our work looks specifically at the answers for which such problems have already been pointed out in the form of comments, and based on which, the answer has been updated to fix the problem.

%%moved to general citation
Zhang et al.~\cite{EmpricalStudyOfObsoleteAnswers} studied obsolete answers on Stack Overflow by analyzing answer comments. 
%Their strategy for determining whether an answer was obsolete or not was based on two criteria:
%\begin{enumerate}
%	\item A comment contained a key phrase of "deprecated", "outdated", "obsolete", or "out of date"
%	\item The key phrase that was observed in criteria 1 does not appear in the question. If the key phrase does appear in the question itself then the question was already asking for an obsolete topic.
%\end{enumerate}
They found that most obsolete answers were already obsolete when they were first posted, and that most reactions to an obsolete answer happened an average of 118 days after the obsoleteness was even observed.
They also found that most answers are not updated when observed to be obsolete and that there are certain languages that are more prone to obsolete answers than others, particularly the languages that are related to mobile application development.
While they focused specifically on answers that were deemed obsolete, our study considers all forms of improvements and code edits, including errors in the code, non-functional improvements, and extensions.

\paragraph{Clarification Comments.} Rao et al.~\cite{RaoDaumClarificationQuestions} used a neural network to learn different kinds of clarification questions that were asked in the question comments to improve the question, e.g., What version of X are you using?
While they do perform some matching of the comments posted on a question to the question edits, they focus only on explicit question statements found in comments (i.e., a sentence that ends with a question mark).
They also did not compare the content of the comment to that of the edit, and assume that the first edit after a question is posted in a comment is the response to that question. 
Along similar lines, Jin et al.~\cite{JinEditsOnHighlyAnsweredQuestions} studied how edits to a question affect the answers the question receives. 
%Although they analyzed the kinds of edits that were made, these edits were made to a question for the purpose of getting more or better answers.
They focused on the edits made to a question before and after it received an accepted answer and how these edits affect the quality of received answers. 
In contrast to both efforts, we try to match code terms in a comment and an edit, and we focus on answer edits rather than question edits.

\paragraph{Summary}  Apart from various technical/methodological differences noted above, the most important differences to prior work on Stack Overflow data are (1) the motivation of our work for constructing data sets that have before/after code versions with associated explanations, (2) we analyze the contents of both comments and edits in order to match them, (3) we extract pairs of comments and their corresponding edits, (4) we consider all types of changes and do not pre-limit ourselves to one type of edit, and (5) we study various characteristics, such as tangledness and usefulness, of these comment-edit pairs.

%While Diamantopoulos et al. focused solely on Java answers and used a neural network to learn about the edits made, we focused on the 5 tags with the most edits: Java, Javascript, Android, Php, and Python and used the comments on an answer to determine edit causation.
%Both Diamantopoulos et al.~\cite{Diamantopoulos:2019:TMA:3341883.3341922} and us used the SOTorrent dataset and used the \textit{PostBlockVersion} to separate the text and code blocks.
%Diamantopoulos et al.~\cite{Diamantopoulos:2019:TMA:3341883.3341922} compared both the text and code blocks of an answer with the changes made to those blocks by subsequent edits to that answer.
%After clustering the results they were able to see the kind of edits were common and were able to recommend edits given a new answer.
%%(The \textit{Comment} in their dataset is the comment attached to an edit and not the comments on the answer).
%In contrast we did not look at the text blocks of an answer and focused only on the code blocks.
%We analyzed the changes between a code block and the changes introduced by an edit with the comments made on that answer to see if we would be able to match a comment with having caused the change in the code block.
%Afterward categorizing the causing comments to see what most answers needed editing for.
\section{Mapping Comments to Edits}
\label{sec:mappingmethod}

In this section, we describe our method for matching comments to edits.
Our goal is to extract comment-edit pairs 
$(c_i, e_j)$, where comment $c_i$ caused edit $e_j$ to occur.

As our main data source, we use the SOTorrent data set~\cite{BaltesSOTorrent} which captures the edit history of all Stack Overflow posts (we use version 2019-09-23).
In SOTorrent, a Stack Overflow post is split into text and code blocks, based on the html formatting used in the post.
\textit{Text blocks} mark any text in the post, including inline code, while \textit{code blocks} mark explicit code blocks formatted using the \texttt{<code>} html tag or the markdown back-tick symbol.  
%An example from the original SOTorrent paper is provided in Figure \ref{fig:TextBlockCodeBlockExample}.
%
%\begin{figure}[t!]
%  \includegraphics[width=0.5\textwidth, scale=0.5]{images/TextBlockCodeBlockExample.png}
%  \caption{Example of an answer being split into text and code blocks. Inline code is still counted as a text block. Source: Adapted from \cite{BaltesSOTorrent}}
%  \label{fig:TextBlockCodeBlockExample}
%\end{figure}
An \textit{edit} to a given post is thus any change to one or more of its text or code blocks.
Given the goal of our work, we focus on edits to code blocks in Stack Overflow answers.
We analyze all answer edits from five popular tags on Stack Overflow: Java, JavaScript, PHP, Python, and Android.
We choose these tags because, at the time of writing, they had the highest number of answers on Stack Overflow.
%These tags had the most answers on Stack Overflow at the time of writing.
The five tags contain a total of \checkNum{11,119,517} answers, \checkNum{12,130,068} comments, and \checkNum{4,322,506} edits.

\subsection{Ground Truth Creation}
\label{sec:groundtruth}

As a first step, we create a ground truth that can help us evaluate and refine any automated matching technique we develop.
To select the answers that we will include in our ground truth, we use stratified sampling to select 20 answers from each tag.
Our stratification strategy selects two answers in each of the following categories: high (above 1000) score, low (below zero) score, recent creation date (after Jan 01, 2018), and old creation date (before Jan 01, 2009). 
Our intuition behind this stratified sampling is to ensure the diversity of answers we examine.
Since answer score is a commonly used metric for answer quality, we want to select answers with extreme scores.
Similarly, we want to select answers from the beginning of Stack Overflow (2008) and recent answers from Stack Overflow (2018) to ensure that we see answers with diverse history.
This resulted in eight selected answers.
We then consider two factors to sample additional answers: (1) the number of comments and (2) the number of edits; these two factors may have direct impact on an automated matching technique so we again want to ensure diversity in our selection.
For each of these factors, we consider two levels: (a) large (more than 10) and (b) small (less than 10).
We sample two answers from each of the four combinations of these factors and levels (i.e., two answers with more than 10 comments and more than 10 edits, two answers with more than 10 comments and less than 10 edits, etc).
This results in eight more answers.
The intention of using 10 as the threshold for a ``large" and ``small" is because we find that the majority of answers have less than 10 edits and less than 10 comments. For the goal of diversifying the sample, we also select answers that have more than 10 edits and/or comments.
Finally, we select four additional random answers with at least one edit and one comment to create our 20 answers for each tag.
In total, our ground truth contains 100 answers with a total of \checkNum{521} edits and \checkNum{799} comments.

%\begin{figure}[t!]
%  \includegraphics[width=0.5\textwidth, scale=0.45]{images/GroundTruthBubblePlot.png}
%  \caption{Distribution of answers in the ground truth, w.r.t the number of comments and edits for each answer}
%  \label{fig:GroundTruthBubblePlot}
%\end{figure}

%Figure \ref{fig:GroundTruthBubblePlot} shows the distribution of the number of answers in relation to the number of comments and edits in the ground truth.
%As shown in the figure, the selected ground truth sample covers a wide range of answers, with varying number of comments and edits.

The two authors then independently evaluated all 100 answers.
For each comment on an answer, they separately analyze the edits for each answer to determine if the comment caused an edit using the following criteria:
\begin{enumerate}
\item The edit occurred after the comment. 
\item The topic of the comment is related to the update in the edit.%, and is the causing reason for the edit.
\end{enumerate}

We use only the above criteria to mark a comment as having caused an edit; it did not matter if the edit affected a text block or a code block or if the comment contained any code.
This was intentional to avoid any bias towards our heuristics of using code terms for matching comments to edits, which we describe later in Section~\ref{sec:matching}. 
For example, in Answer\footlink{http://stackoverflow.com/questions/281433} 281433, we manually match the comment \textit{``But he is not calculating a simple mean. Remember there were only three votes given in his example.''} to Edit\footlink{https://stackoverflow.com/revisions/281433/3} 3 that removed the SQL query that implemented a simple mean, even though there are no explicit code terms used in the comment.
%To ensure that our coding guide is clear, the coders compared results on the first few answers from Java and Javascript and discussed disagreements.
%The causes for disagreements mainly came from cases where multiple comments were discussing the change.
%For example, comment \texttt{c1} in Figure~\ref{fig:55224131_example} from \post{55224131} was marked by one author as having caused Edit 4 which changed the line \texttt{exec('start /launch\_vbs.bat');} to \texttt{exec('start launch\_vbs.bat');} on March 18, 2019 15:33:30, while the second author marked \texttt{c7} as causing the edit. 
%The first author's reasoning was that \texttt{c1} started the conversation that resulted in the edit, and thus is the cause for the edit, while the second author's reasoning was that the actual error that helped in determining what to edit was pointed out in \texttt{7}.
%% It is edit 4
%To resolve this, both authors agreed on the following rule, in addition to the above coding guide: in the case of a conversation (i.e., multiple comments) that discussed an issue that ends up resulting in an edit, they would mark all the comments (\checkNum{where the comment author is not same as the edit author}) as relating to the edit.
The two authors then discussed and resolved any disagreements.
For any labelling/coding exercise throughout this paper, we resolved disagreements as follows:
together, the two authors discuss each disagreement and justify their label for the comment-edit pair in question.
The authors continue discussing the pair until an agreement is reached.
%If an agreement was not made between the authors then it was agreed to default on the side of precision.
Creating this ground truth set took around 26 hours, as both authors need to analyze \textit{all} comments and edits for each selected answer. 
Overall, our Cohen's Kappa score~\cite{mchugh2012interrater} 
 for matching comment-edit pairs is \checkNum{0.71}.
%For comment-edit pairs that both coders agreed on, they also then marked whether the comment-edit pair was useful.
%As mentioned in the introduction, we consider a pair as \textit{useful} if (1) the edit happens to an existing code snippet in the answer such that there is code to potentially be matched in a target system and (2) if the comment describes this change in a way that is understandable outside of the posed Stack Overflow question.
%This means that the edits must be modifications or additions to existing code snippets in the answer, as opposed to text additions or completely new code snippets.
%For example, if the edit added an example that did not previously exist, then that comment-edit pair would be considered \textbf{not} \textit{useful}, but if an edit added details to a code snippet and did not remove anything, that \textbf{would} be considered \textit{useful})
%The two authors marked the usefulness of comment-edit pairs as either: \textit{Yes}, \textit{No}, or \textit{Yes with edits}.
%Yes with edits means that the comment-edit pair is useful, but the comment needs some editing. i.e., to make the comment more understandable, or to remove conversational ``fluff'' such as expressing thanks.
%we will be tight in space so removed this figure
%\begin{figure*}[t!]
%  \includegraphics[width=\textwidth, scale=0.5]{images/281433_Example.png}
%  \caption{The two comments on the left have each caused an edit even though they did not suggest any code
%  }
%  \label{fig:281433_Example.png}
%\end{figure*}

Table~\ref{tab:GroundTruthStatistics} shows the descriptive statistics per tag in our ground truth.
In total, we analyzed 100 answers with \checkNum{799} comments and \checkNum{521} edits to construct a ground truth of \checkNum{194} comment-edit pairs.

\begin{table}[t!]
\centering
\caption{Ground Truth Statistics}
\resizebox{\textwidth}{!}{
\begin{tabular}{@{}lrrrrrrrl@{}}
\toprule
\multicolumn{1}{m{0.5cm}}{\textbf{Tag}} 
& \multicolumn{1}{m{1.2cm}}{\textbf{Answers}} 
& \multicolumn{1}{m{.75cm}}{\textbf{Edits}} 
& \multicolumn{1}{m{1.6cm}}{\textbf{Comments}} 
& \multicolumn{1}{m{1.8cm}}{\textbf{Median comments}} 
& \multicolumn{1}{m{1.3cm}}{\textbf{Median edits}} 
& \multicolumn{1}{m{2cm}}{\textbf{Comment-edit pairs}} \\
%& \multicolumn{1}{m{1cm}}{\textbf{Kappa Score}} 
%& \multicolumn{1}{m{1.5cm}}{\textbf{Precision}} 
%& \multicolumn{1}{m{0.5cm}}{\textbf{Recall}} \\
\midrule
\textbf{Java}       & 20 & 95 & 148 & 5.5 & 2.0 & 38\\ %& \checkNum{85\%} & \checkNum{47\%} & \\
\textbf{JavaScript} & 20 & 105 & 158 & 6.0 & 3.0 & 33 \\ %& \checkNum{71\%} & \checkNum{30\%} & \\
\textbf{Android}    & 20 & 101 & 202 & 8.5 & 3.0 & 40 \\ %& \checkNum{56\%} & \checkNum{36\%} & \\
\textbf{Python}     & 20 & 103 & 136 & 5.5 & 3.0 & 38 \\ %& \checkNum{69\%} & \checkNum{23\%} & \\
\textbf{Php}        & 20 & 117 & 155 & 6.0 & 3.0 & 45 \\ %& \checkNum{69\%} & \checkNum{24\%}   \\
\midrule
\midrule
\textbf{Total}      & 100 & 521 & 799 & - & - & 194  \\ %& - & - \\
%\textbf{Average}    & - & - & - & - & - & - & \checkNum{70\%} & \checkNum{32\%}\\
\bottomrule
\end{tabular}
}
\label{tab:GroundTruthStatistics}
\end{table}

\subsection{Automatically Matching Comments and Edits}
\label{sec:matching}

%\begin{figure}[t!]
%  \includegraphics[width=0.5\textwidth, scale=0.5]{images/55224131_example.pdf}
%  \caption{Comment $C1$ starts the conversation about \texttt{exec} but comment $C7$ contains the actual problem}
%  \label{fig:55224131_example}
%\end{figure}

\paragraph{Algorithm Overview.} Given our motivation that mined comment-edit pairs can be later used for creating \maintenance data sets for use in various recommender systems, we only consider edits to code snippets. 
Based on that, the high-level idea of the algorithm is that if a comment mentions a code term that then gets removed or added in a later code edit, we can reasonably assume that the comment caused that edit.
Following the analysis of the 100 ground truth answers, we also add the criterion that the comment-edit pairs are considered only if the users are different.
This is because during the manual analysis, we noticed that when the users are the same, it was difficult to be certain that their own comment \textit{caused} the edit.
It could be the case the user was originally intending on making an edit and first commented an explanation.
Thus, for the sake of precision, we add this criterion to our automated analysis.

\paragraph{Data Preparation.} As a first step, we create two tables that are necessary to store the post-processed SOTorrent data that is relevant for our analysis.
The first table we construct is adapted from the \texttt{EditHistory} table based on a blog post from Baltes~\cite{BaltesEditHistory}, one of the authors of the SOTorrent data set.
This table keeps track of questions, answers, comments, and edits to both the questions and answers.
This table also provides the creation date for each of these events and allows us to order the edits and comments in chronological order.
We include the parent post ID in this table to allow us to find all the answers, edits, and comments related to a specific question.
The second table we create is called \texttt{EditHistory\_Code}, which is built from the \texttt{EditHistory} table and is similar except that instead of containing all changes in the edits, it contains only answers with code blocks and the corresponding edited text from only code edits.
We obtain the actual code edits from the \texttt{PostBlockVersion} table provided in the SOTorrent data set~\cite{BaltesSOTorrent}.
The \texttt{EditHistory\_Code} table we construct contains all the initial body of an answer, its subsequent edits, and comments to the answer in chronological order, while removing all unecessary data such as the title version history and textual answers and edits.
Our program needs only the \texttt{EditHistory\_Code} table to analyze whether comments cause edits to answers.

\begin{lstlisting}[language=Python, caption=Algorithm for matching comments to edits, float, label={lst:match-algorithm}, mathescape=true, escapechar=|,numbers=left,numbersep=5pt,xleftmargin=10pt,basicstyle=\footnotesize,floatplacement=t!,captionpos=b]
matched_pairs = $\emptyset$
for $a_i$ in all_answers: |\label{line:allanswers}|
  for $c_j$ in comments($a_i$):  |\label{line:allcomments}|
    comment_code_terms = extractCodeTerms($c_j$) |\label{line:extractcommentcode}|
    prev_edit = $e_1$ 
    for $e_k$ in edits($a_i$): |\label{line:alledits}|
      if date($e_k$) > date($c_j$) and $c_j$_author != $e_k$_author: |\label{line:comparedate}|
        edit_code_terms = extractCodeTerms($e_k$) |\label{line:extractcurr}|
        prev_edit_code_terms = extractCodeTerms(prev_edit) |\label{line:extractprev}|
        edit_code_diff = edit_code_terms $\bigtriangleup$ prev_edit_code_terms |\label{line:editcodediff}|
        code_matches = edit_code_diff $\cap$ comment_code_terms |\label{line:codematch}|
        if code_matches: |\label{line:checkmatch}|
          matched_pairs = matched_pairs $\cup$ ($c_j$, $e_k$)
          break  |\label{line:endcheckmatch}|
      prev_edit = $e_k$
\end{lstlisting}

\begin{figure*}[t!]
  \includegraphics[width=\textwidth, scale=0.5]{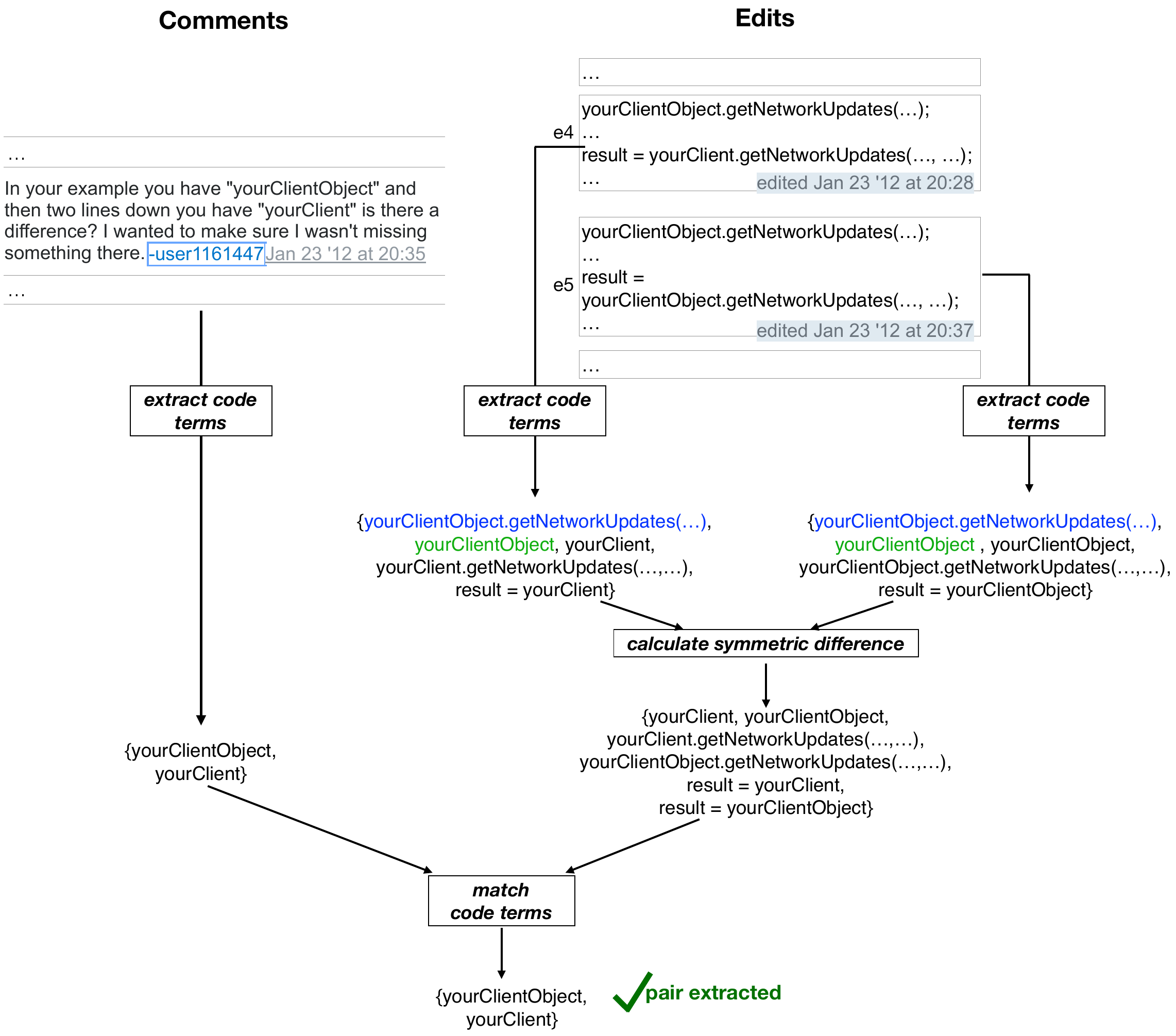}
  \caption{Example from the SO answer 8949391 showing the matching process between comments and edits, based on code terms. The comment shown is matched to edit e5. Example has been reproduced and edited for better visualization. Note that we record a list of code terms, which takes into account how many times a code term appears. In this example, \texttt{yourClientObject} appears twice in the $e5$ code terms.}
  \label{fig:8949391_example}
\end{figure*}

\paragraph{Algorithm Details.} Listing~\ref{lst:match-algorithm} shows the algorithm we use to match comments to edits.
We use the example in Figure~\ref{fig:8949391_example} as a running example to explain the algorithm.
For each answer in the data set (Line~\ref{line:allanswers}), the program iterates through all the comments in chronological order (Line~\ref{line:allcomments}).
It then extracts all code terms found in a comment, storing them in \texttt{comment\_code\_terms} (Line~\ref{line:extractcommentcode}). 
Figure~\ref{fig:8949391_example} shows the extracted comment code terms on the left side of the figure.
To extract code terms, we first look for explicit markdown or html tags (i.e., \texttt{<code>}).
However, not all users strictly follow the formatting guidelines, and comments on Stack Overflow are diverse in the ways they contain code. 
For example, some comments paste code from the answer that did not work for them, while others post comments on the exception that occurred for them.
Some users use the markdown code symbol while others do not and instead paste the code as plaintext.
To simplify the task of extracting code terms, we use regular expression patterns that identify code terms and do not rely solely on markers or formatting guidelines.
Our regular expressions therefore catch code terms by, for example, matching camel case or snake case identifiers, or matching method calls.
%Since not all users strictly follow the formatting guidelines, we also use a series of regular expression patterns to detect code terms that have not been explicitly formatted.
We start with the list of regular expressions used by Treude et al.~\cite{TreudeRegex}.
We modify some of the expressions based on testing on the ground truth set and also remove unnecessary or problematic expressions.
Since the original set of expressions was developed mainly for Java, we also add additional regular expressions catered to the other languages in our data set.

To illustrate our use of regular expressions, we use the following two examples of Stack Overflow comments that contain different formats/styles of code terms: (1) \textit{``The question doesn't mention the user entering *EXIT*. Also, System.exit(0) will terminate the whole JVM, which means that all processing done by the code till that statement will be lost."} on Answer\footlink{http://stackoverflow.com/questions/52347606} 52347606 and (2) \textit{``Sorry, I'm coming to this late, but shouldn't \`{}vars(a)\`{} do this? For me it's preferable to invoking the \`{}\_\_dict\_\_\`{} directly."} on Answer\footlink{http://stackoverflow.com/questions/62680} 62680. Notice that the first example comment does not have code formatted with any explicit code formatting tags, while the second one does. Our corresponding regular expressions that identified the code terms in these two comments, in respective order, are \textit{[a-zA-Z0-9.\_()'\#\$\textbackslash "]+\textbackslash (.*\textbackslash )+}, which matches method calls with dot accesses, and \textit{\_\_[\^\_]*\_\_}, which matches everything between two underscores on either side. 
The full list of regular expressions we use can be found in our artifact page~\cite{artifact}.

The algorithm then iterates over all edits for this answer, in chronological order, to try to match them to the current comment (Line~\ref{line:alledits}).
When the program finds an edit that was made after the comment (Line~\ref{line:comparedate}), it extracts the code terms found in the current edit (which has the snapshot of the code after the change) and the previous edit (which has the snapshot of the code before the change), using the same code identification technique used for comments (Lines~\ref{line:extractcurr}-\ref{line:extractprev}).
The program then takes the symmetric difference between these two lists of code terms to determine any added or removed code terms (Line~\ref{line:editcodediff}).
In Figure~\ref{fig:8949391_example}, the symmetric difference of the edits is displayed on the right side of the figure.
The common code terms between between the current edit and the previous edit are shown in the same color.
The symmetric difference contains all the remaining terms, which appear only in one of the edits.
Finally, our algorithm compares the code terms found in the comment to the code terms found in the symmetric difference between the two edits (Line~\ref{line:codematch}).
Since the code term used in the comment may not be exactly the same as that used in the code due to typos or placeholder text in the code snippet, we calculate the Levenshtein distance~\cite{Levenshtein_SPD66}, using the \textit{fuzzywuzzy} library in Python~\cite{fuzzywuzzy}, between the code terms in the comments and those in symmetric difference to determine a match. 
We consider two code terms as a match if their similarity ratio is above 90\%.

\begin{figure*}[t!]
\centering
\includegraphics[width=0.8\textwidth]{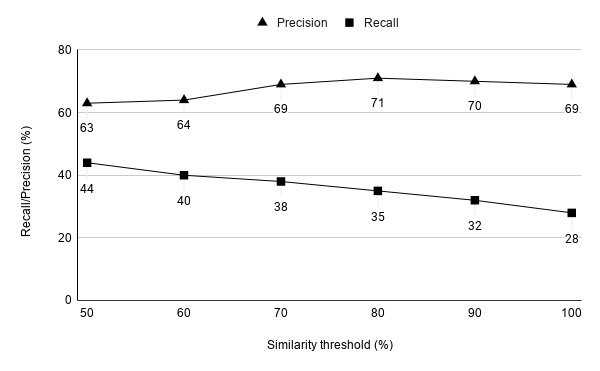}
\caption{Comparison of similarity threshold used to identify matching code terms}
\label{fig:threshold_comparison}
\end{figure*}

\label{ref:threshold-choice}
We choose the 90\% threshold based on examining the results of varying thresholds. According to Figure~\ref{fig:threshold_comparison}, we can see that at an 80\% threshold results in the highest precision.
However, what is not conveyed through this graph are the number of terms that are caught by the program at the various thresholds.
With the original goal of having comments as explanations for edits, we want to as accurately as possible select the code terms in the comment that get edited in the answer.
Since precision reflects the percentage of matched comment-edit pairs and not which code terms get matched, the precision between the different thresholds does not change significantly.
In other words, matching one code term is the same as matching five code terms; in both cases, the comment and edit will be matched.
When we manually analyzed the matched code terms made by the program at the 80\% and 90\% thresholds, we found that using the 90\% threshold removes some code terms that are caught at the 80\% threshold but do not contribute to the edit. For example, in Answer\footlink{http://stackoverflow.com/questions/34459380} 34459380, the comment that causes Edit\footlink{https://stackoverflow.com/revisions/34459380/3} 3 is: \textit{``So in this example is theArray also the key in the local storage. so for me if I had the key as keyword and the array as myArray would it then be, localStorage.setItem(`keyword', JSON.stringify(myArray)); ?"}
The matched edit contains the addition of example functions. 
One change adds \textit{``function setArrayInLocalStorage(key, array) \{ localStorage.setItem(key, JSON.stringify(array));\}"}.
In this example, our program at an 80\% threshold matches the following terms between the comment and the edit \textit{``[`setItem', `localStorage', ``localStorage.setItem('keyword', JSON.stringify(myArray))",\\  ``setItem('keyword', JSON.stringify(myArray))", `theArray']"}, while at the 90\% threshold, it returns \textit{``[`localStorage', `setItem', `theArray']."}
%For example, in answer \post{43673662} the comment that caused \edit{https://stackoverflow.com/posts/43673662/revisions}{edit 2} is: \textit{``holder.cardView.setOnClickListener(new View.OnClickListener() \{ @Override public void onClick(View v) \{ Toast.makeText(holder.itemView.getContext(), ``cardview", Toast.LENGTH\_SHORT).show(); \} \}); no luck "} and the relevant change in the edit is the replacement of \textit{``holder.itemView.getContext()"} with \textit{``mContext"} which is an added \textit{``private Context mContext;"} field that is instantiated in the constructor \textit{``this.mContext = \_context;"}.
%In this example, our program at an 80\% threshold matches \textit{``[`Toast.makeText(holder.itemView.getContext(), ``cardview", `Toast.LENGTH\_SHORT).show()', `makeText(holder.itemView.getContext(), ``cardview", Toast.LENGTH\_SHORT).show()', `getContext', `itemView']"} in the comment, while at the 90\% threshold it returned \textit{``[`getContext', `itemView']"}.
From this example, with the original goal in mind, it is preferable to have the returned matches of the 90\% threshold rather than the 80\% threshold as it provides a more accurate depiction of which code terms in the comment truly attributed to the changes in the edit.
While the difference in precision and recall is insignificant between the 80\% and 90\% thresholds, the previous preference of accuracy of the related code terms explains why we chose the 90\% threshold. % rather than an 80\% threshold.
More details on the difference between the thresholds can be found on our artifact~\cite{artifact} page.

If the program finds a match between a code term in the comment and a code term in the edit, it labels the comment as having resulted in the edit, and adds this comment-edit pair to the set of matched pairs  (Lines~\ref{line:checkmatch}-~\ref{line:endcheckmatch}). 
In Figure~\ref{fig:8949391_example}, the matched code terms (\texttt{yourClient} and \texttt{yourClientObject}) are shown at the bottom of the figure.
Since there are matched code terms between the comment and the edit, in this example, we would say that the given comment is matched with \texttt{e5}.
Note that the \texttt{break} on Line~\ref{line:endcheckmatch} indicates that a comment is matched to the first edit it is related to.

%Equation~\ref{LevenshteinDefinition} shows how the Levenshtein distance for two terms is calculated as a ratio.
%Where $a, b$ are two strings and $i, j$ are the starting indices for comparison.

%\begin{equation}
%lev_{a,b}(i,j) = 
%\begin{cases}
%	max(i,j)
%	&\text{if min(i,j) = 0,} \\
%	min
%	\begin{cases}
%	lev_{a,b}(i-1,j) + 1 \\
%	lev_{a,b}(i,j - 1) + 1 \\
%	lev_{a,b}(i - 1,j - 1) + 1_{a_i \neq b_j} \\
%	\end{cases} 
%	&\text{otherwise.}
%\end{cases}
%\label{LevenshteinDefinition}
%\end{equation}
%\begin{equation}
%\frac{(|a| + |b|) - lev_{a,b}(i,j)}{|a| + |b|}
%\label{SimilarityRatio}
%\end{equation}

\subsection{Comparison with Ground Truth}
\label{sec:groundtrutheval}

Before running our automated matching strategy on all the data we have for all tags, we want to evaluate its effectiveness and fix any issues. 
Thus, we run the above matching algorithm on the manually created ground truth set of 100 answers from Section~\ref{sec:groundtruth} and calculate recall and precision.
Recall is the percentage of comment-edit pairs the program could detect from the manually confirmed pairs in the ground truth, while precision is the percentage of comment-edit pairs identified by the program that are correct.
Additionally, to understand if the code matching algorithm we use brings in any value, we compare our results to those of a simple baseline.
This baseline simply matches a comment to the chronologically nearest edit that comes after it, regardless of the content of the comment or edit.
We show the results in Table~\ref{tab:GroundTruthComparison}.
As shown, the recall of our matching technique is low~(\checkNum{ranging from 24\% - 47\% and 32\% overall}), but the precision is relatively good (ranging from \checkNum{56\% - 85\%} and \checkNum{70\%} overall. 
To understand when our matching fails, we manually analyze the false positives and false negatives.

\begin{table}[t]
\centering
\caption{Matching Evaluation on Ground Truth Data Set}
\label{tab:GroundTruthComparison}
\resizebox{\textwidth}{!}{%
\begin{tabular}{@{}rrrrrrrr@{}}
\toprule
\multirow{2}{*}{\textbf{Tag}} & 
\multirow{2}{*}{{\textbf{\begin{tabular}[c]{@{}c@{}}Existing\\ Pairs\end{tabular}}} } & 
\multicolumn{3}{c}{\textbf{Our Matching Program}} & 
\multicolumn{3}{c}{\textbf{Proximity Based Baseline}}\\ 
\cmidrule(l){3-5} \cmidrule(l){6-8}
                     &                                 & Detected & Recall & Precision  & Detected & Recall & Precision  \\
\midrule
\textbf{Java} & 38 & 20 & 47\% & 85\%  & 81 & 64\% & 28\%  \\
\textbf{JavaScript} & 33 & 14 & 30\% & 71\%  & 65 & 70\% & 35\%  \\
\textbf{Android} & 40 & 25 & 36\% & 56\%  & 96 & 69\% & 28\% \\
\textbf{Python} & 38 & 13 & 23\% & 69\%  & 59 & 53\% & 34\%  \\
\textbf{Php} & 45 & 16 & 24\% & 69\% & 63 & 51\% & 37\%  \\
\midrule
\textbf{Overall} & 194 & 88 & 32\% & 70\% & 364 & 60\% & 32\%  \\
\bottomrule
\end{tabular}%
}
\end{table}

%%with usefulness
%\begin{table*}[]
%\centering
%\caption{Matching Evaluation on Ground Truth Data Set}
%\label{tab:GroundTruthComparison}
%\resizebox{0.7\textwidth}{!}{%
%\begin{tabular}{@{}llllllllll@{}}
%\toprule
%\multirow{2}{*}{\textbf{Tag}} & 
%\multirow{2}{*}{{\textbf{\begin{tabular}[c]{@{}c@{}}Existing\\ Pairs\end{tabular}}} } & 
%\multicolumn{4}{c}{\textbf{Our Matching Program}} & 
%\multicolumn{4}{c}{\textbf{Proximity Based Baseline}}\\ 
%\cmidrule(l){3-6} \cmidrule(l){7-10}
%                     &                                 & Detected & Recall & Precision & Useful & Detected & Recall & Precision & Useful \\
%\midrule
%\textbf{Java} & 38 & 23 & 47\% & 85\% & 82\% & 81 & 64\% & 28\% & 74\% \\
%\textbf{Javascript} & 33 & 14 & 30\% & 71\% & 40\% & 65 & 70\% & 35\% & 39\% \\
%\textbf{Android} & 40 & 25 & 36\% & 56\% & 29\% & 96 & 69\% & 28\% & 33\% \\
%\textbf{Python} & 38 & 13 & 23\% & 69\% & 33\% & 59 & 53\% & 34\% & 35\% \\
%\textbf{Php} & 45 & 16 & 24\% & 69\% & 45\% & 63 & 51\% & 37\% & 35\% \\
%\midrule
%\textbf{Total} & 194 & 88 & 32\% & 70\% & 48\% & 364 & 60\% & 32\% & 41\% \\
%\bottomrule
%\end{tabular}%
%}
%\end{table*}

\label{page:lowrecallexpl}
One of the main reasons for the low recall (i.e., false negatives) is that there are comments in the ground truth that caused an edit but did not contain any code suggestions. 
Our program is only able to pair comments and edits that share a code pattern; as such it is not able to find these comment-edit pairs.
An example of this is Answer\footlink{http://stackoverflow.com/questions/44765572} 44765572.
Here, both authors agreed that the comment on Jun 26 '17 at 18:13: \textit{``I think it would be a lot cleaner to have the constructor accept the three parameters, instead of always creating it with the defaults and then overwriting them."} caused Edit\footlink{https://stackoverflow.com/revisions/44765572/9} 9, that adds the parameters to the constructor instead of overwriting the default values.
%This comment states a preference that improves the code.
%For this particular context, if the default values are always overwritten then it would be cleaner if the class required the necessary values before instantiation.
% This would be an example of \textit{preventative maintenance}.
%This example illustrates one of the scenarios that contribute to the main reason of low recall of the program.
This comment does not use explicit code terms to detail inefficiencies or problems but rather explains how the code can be improved.
%The edit is then made to incorporate the suggestion but there is no code terms in the comment for our program to compare to the edit.
Other comments that cause edits without having explicit code could be questions clarifying the logic of the answer, or comments mentioning the answer does not fully answer the original question, etc.
All of these comments would cause edits but would not have any code for the program to match the comment and edit together.

While our program, by construction, is not able to capture such pairs and it might have been more ``fair'' to evaluate our program only on the comment-edit pairs it could potentially capture (i.e., those with code), we chose to conduct a strict evaluation to understand the worst case performance of the algorithm in terms of how many pairs it could potentially capture.
However, for further investigation into recall, we also check what our program would have done only on pairs it could potentially match.
While annotating the ground truth set of 799 comments, the two authors agreed on 194 comments that caused an answer edit (See Table~\ref{tab:GroundTruthStatistics}).
Out of these 194 comments, 135 comments had code terms that the program could potentially match. This means that 59 (30\%) of the pairs can not, by construction, be found by the program as there is no code for it to match.
If we evaluate the matching algorithm on only the 135 comments that it could potentially match, we find that we still maintain a 69\% overall precision, but now achieve a 46\% overall recall instead of 32\%.
This confirms our intuition that many of the false negatives are due to the comments not having any code to match.
Other reasons for the remaining false negatives include potentially missed regular expressions for detecting code terms, matching the wrong edit if it contains the right code term and happens earlier than the ground truth edit, or that the answer edit itself is an added textual explanation rather than an edit to a code snippet.

On the other hand, the majority of false positives occur, because of coincidental matches between a comment and an edit,
i.e., the program finds a code suggestion both in a comment and an edit, but the edit was not caused by that comment.
An example for this is Answer\footlink{https://stackoverflow.com/questions/6872517} 6872517.
Our program detects that comment \textit{``thanks , as i see on findViewById(R.id.mainframe) , i need to add a id ? and a layout file ?"} caused Edit\footlink{https://stackoverflow.com/revisions/6872517/2} 2. However, this edit simply properly formats the whole code snippet without addressing this comment in any way.
The program catches this edit and matches the code in the comment (\texttt{findViewById(R.id.mainframe)}) to the code in the edit and assumes a relationship when there is none.

While in an ideal world, an automated technique would have both high recall and precision, in practice, there is often a tradeoff between both metrics.
For the purposes of using the extracted pairs to build data sets, we believe it is more important to have high precision than high recall.
Given the vast amount of data available on Stack Overflow, extracting even a tiny fraction of available comment-edit pairs will provide a large amount of data.
However, if this data contains a large number of false positives, then its users will lose their trust in the data.
Thus, it is important for the matching technique to have high precision, even if this is at the cost of missing out on other potential pairs.
We, do, however, discuss opportunities for improving recall in Section~\ref{sec:discussion}.
When compared to the proximity based baseline, our program achieves a much higher overall precision (70\% vs. 32\%), which gives us confidence in using our matching algorithm to answer our five research questions.
%This give us confidence that the comment-edit pairs our program extracts are generally correct.

\section{\ref{rq:precision}: Precision of Comment-Edit Pairs} 
\label{sec:rq1}

We now discuss \ref{rq:precision}, which focuses on the precision of our automated mapping strategy.
While the ground truth evaluation gave us confidence to proceed, our ground truth is still limited in size. Thus, for \ref{rq:precision}, we run our matching program on the data from all five tags. We first describe our evaluation methods and then report the results.

\paragraph{Methods.}

We first run our matching program on the data from all five tags we focus on.
Table~\ref{tab:LanguageStatistics} shows the descriptive statistics for this data, as well as the number of comment-edit pairs detected by our tool.

\begin{table}[t!]
\centering
\caption{Number of answers, edits, and comments in each of the five Stack Overflow tags, as well as the number of comment-edit pairs we detect for each tag}
\resizebox{0.75\textwidth}{!}{
\begin{tabular}{@{}lrrrr@{}}
\toprule
\textbf{Language}   & 
\multicolumn{1}{l}{\textbf{Answers}} & 
\multicolumn{1}{l}{\textbf{Edits}} & 
\multicolumn{1}{l}{\textbf{Comments}} & 
\multicolumn{1}{m{1.5cm}}{\textbf{Detected comment-edit pairs}} \\
\midrule
\textbf{Java}       & 2,586,447 & 895,737  & 2,321,296 & 51,358 \\
\textbf{JavaScript} & 2,924,662 & 1,281,433 & 3,571,622 & 65,373 \\
\textbf{Android}    & 1,722,580  & 490,565  & 1,668,634 & 34,596 \\
\textbf{Python}     & 1,785,914 & 903,159  & 2,060,513 & 44,551 \\
\textbf{Php}        & 2,099,914 & 751,612  & 2,508,003 & 52,521 \\
\bottomrule
\end{tabular}
}
\label{tab:LanguageStatistics}
\end{table}

Calculating precision requires manually analyzing the detected pairs.
Since it is not feasible to manually validate close to 250,000 pairs, we take a statistically representative sample for each tag.
For a confidence level of 95\% with a 5\% confidence interval, we need a sample size of 382 pairs for each tag.
Therefore, we randomly select 382 pairs from each tag for our manual validation, resulting in a total of \checkNum{1,910} comment-edit pairs to be validated.

The two authors of the paper then separately analyze all \checkNum{1,910} comment-edit pairs, with the goal of confirming whether the identified comment is related to the corresponding edit in the pair.
Determining if a comment-edit pair is correct and gathering additional data about its usefulness, category, and tangled changes takes on average 1.5 minutes.
Thus, the two authors spent close to 95 hours to manually analyze the \checkNum{1,910} pairs.
An additional \checkNum{8hrs} (approximately 1.5 hours per tag) were taken to resolve conflicts since conflict resolution involved more discussion.

After the resolutions, each comment-edit pair was labelled with either zero (comment is not related to the edit) or one (comment is related to the edit).
We use Cohen's Kappa score~\cite{viera2005understanding} to calculate the inter-rater agreement rate.

\paragraph{Results.} Table~\ref{tab:precision} shows the precision of our matching strategy, as well as Cohen's Kappa, for each analyzed Stack Overflow tag. 
The last row of the table shows the overall aggregate results over all analyzed data.

As shown, our Kappa score ranged \checkNum{0.67-0.86} across the five tags. Out of the 1,910 pairs we analyze, we confirm \checkNum{1,482} pairs. The precision per tag ranges from \checkNum{74-80\%}. When considering all 1,910 pairs, the overall precision of our algorithm is \checkNum{78\%}. We also note that the precision across the five tags is fairly similar, which suggests that our matching heuristics are not biased toward a particular programming language or lexicographical pattern. 

\begin{findingenv}{\ref{rq:precision}}{rq1}
Across the five tags, the precision of our automated comment-edit mapping algorithm is \checkNum{78\%}.
\end{findingenv}

\begin{table}[t!]
\centering
\caption{Precision of Detected Comment-edit Pairs Across the Full Data Set}
\resizebox{0.9\textwidth}{!}{
\begin{tabular}{@{}lrrrr@{}}
\toprule
\textbf{Tag}   & 
\multicolumn{1}{l}{\textbf{Pairs Analyzed}} & 
\multicolumn{1}{l}{\textbf{Pairs Confirmed}} & 
\multicolumn{1}{l}{\textbf{Cohen's Kappa}} & 
\multicolumn{1}{l}{\textbf{Precision}} \\
\midrule
\textbf{Java}       & 382 & \checkNum{305} & \checkNum{0.67} &\checkNum{80}\%\\
\textbf{JavaScript} & 382 & \checkNum{307} & \checkNum{0.77} &\checkNum{80}\%\\
\textbf{Android}    & 382 & \checkNum{284} & \checkNum{0.86} &\checkNum{74}\%\\
\textbf{Python}     & 382 & \checkNum{292} & \checkNum{0.75} &\checkNum{76}\%\\
\textbf{Php}        & 382 &  \checkNum{294} & \checkNum{0.77} &\checkNum{77}\%\\
\midrule
\textbf{Total}      & 1,910 & \checkNum{1,482} & \checkNum{0.77} &\checkNum{78}\%\\
\bottomrule
\end{tabular}
}
\label{tab:precision}
\end{table}

\section{\ref{rq:tangled}: Tangled Changes}
\label{sec:rq2}

Recall that the term \textit{tangled change} refers to grouping separate code changes in a single commit or edit~\cite{herzig2013impact}.
In the introduction, we speculated that one of the attractive qualities of using Stack Overflow edits is that changes on Stack Overflow are likely to be less tangled than those found in commits in version-control systems.
In this research question, we investigate if this is true in practice.

\paragraph{Methods.} 
\label{par:tangled-eg}
For each of the \checkNum{1,482} confirmed comment-edit pairs found in \ref{rq:precision}, we also record whether the edit contains tangled changes or not.
The two authors again independently labeled tangled changes and discussed disagreements.

In the context of comment-edit pairs, tangled changes occur if the edited answer contains additional changes that are \textit{not} related to the matched comment.
An example of a tangled change would be an edit that addresses multiple comments at a time.
For example, in Answer\footlink{https://stackoverflow.com/questions/5616616} 5616616, the original questioner puts the following comment \textit{``Can I add a variable to the id like} $<id= \$count.frDocViewer>$ \textit{and then it would access} $\#\$count.frDocViewer?$ \textit{...''}.
The answerer posts a comment in response to this explaining how they can use the suggested variable.
The questioner then posts another comment on a different part of the code \textit{``Should there be an else statement after} $if(fr != old\_element) \{ fr.style.display =$$"block"$ $old\_element.style.display = "hide";$ $old\_element = fr; \}$ \textit{? Why does there have to be "echo" in} $'HideFrame(echo$\\ $\$count)'$ \textit{?}
At this point, the answerer edits the code snippet\footlink{https://stackoverflow.com/revisions/5616616/5} to fix both the redundant echo and the if statement in question. 
However, they also address the initial comment to show how to correctly use the \texttt{count} variable.
Pairing either of these comments with the edit is an example of a comment-edit pair with a tangled change since the edit addresses changes beyond those related to the matched comment. 
Tangled changes also occur when the answerer does not look at their answer for a period of time while other users view the answer and make comments on what, if any, changes they recommend.
The answerer then returns and decides to create one edit to address all the comments received.
Similarly, a tangled edit includes addressing a single comment but also making cosmetic changes, such as variable renames in the code snippet or text reformulation in the answer. 
For example, Edit\footlink{https://stackoverflow.com/revisions/5169321/6} 6 of Answer\footlink{https://stackoverflow.com/questions/5169321} 5169321 addresses multiple issues that were brought up in the comments such as answering clarification questions, or that the answer still does not solve the question, while at the same time formatting the answer for visual clarity.

%Our inter-rater agreement for this labeling was \todo{Z}.

\paragraph{Results.} 
Table~\ref{tab:tangled} shows the number of tangled pairs, both per tag and overall.
As shown, only \checkNum{11\%} of the total confirmed pairs are tangled.
These results coincide with our intuition that since Stack Overflow snippets and answers are typically short, their edits would mostly focus on one issue at a time.
From our general observations, the main reason for tangled changes are when the answerer includes additional refactorings to make the answer more concise or readable while addressing the feedback in the comment.

\begin{findingenv}{\ref{rq:tangled}}{rq2}
Our results confirm our intuition that the code changes in Stack Overflow comment-edit pairs are rarely tangled. Specifically, only \checkNum{11\%} of the \checkNum{1,482} confirmed comment-edit pairs we analyzed contain tangled changes.
\end{findingenv}

\begin{table}[t!]
\centering
\caption{Number of useful pairs and tangled edits in the confirmed comment-edit pairs}
\resizebox{0.85\textwidth}{!}{
\begin{tabular}{@{}lrrrrr@{}}
\toprule
\multirow{2}{*}{\textbf{Tag}}   & 
\multirow{2}{*}{\thead{\textbf{Confirmed}\\\textbf{Pairs}}} & 
\multicolumn{2}{c}{\textbf{Tangled}} & 
\multicolumn{2}{c}{\textbf{Useful}} \\
\cmidrule(l){3-4}\cmidrule(l){5-6}  
&& \textbf{Kappa Score} & \textbf{Count} (\%)& \textbf{Kappa Score}& \textbf{Count} (\%)\\
\midrule
\textbf{Java}       & \checkNum{305} & \checkNum{0.79} &\checkNum{41 (13\%)}&\checkNum{0.79}&\checkNum{67 (22\%)}\\
\textbf{JavaScript} & \checkNum{307} & \checkNum{0.65} &\checkNum{41 (13\%)}&\checkNum{0.70}&\checkNum{91 (30\%)}\\
\textbf{Android}    & \checkNum{284} & \checkNum{0.59} &\checkNum{23 (8\%)}&\checkNum{0.81}&\checkNum{71 (25\%)}\\
\textbf{Python}     &\checkNum{292} & \checkNum{0.61} &\checkNum{29 (10\%)}&\checkNum{0.78}&\checkNum{107 (37\%)}\\
\textbf{Php}        & \checkNum{294} & \checkNum{0.64} & \checkNum{27 (9\%)}&\checkNum{0.62}&\checkNum{60 (20\%)}\\
\midrule
\textbf{Overall}      &\checkNum{1,482} & \checkNum{0.67} &\checkNum{161 (11\%)}&\checkNum{0.74}&\checkNum{396 (27\%)}\\
\bottomrule
\end{tabular}
}
\label{tab:tangled}
\end{table}
\section{\ref{rq:changetype}: Types of Changes in Comment-Edit Pairs}
\label{sec:rq3}

In \ref{rq:changetype}, we look at the types of changes that occur in comment-edit pairs.
Understanding the types of changes helps determine what code maintenance changes, if any, the extracted comment-edit pairs can be useful in.
For example, let us assume that we find that the majority of comment-edit pairs are simply questions where a commenter asks for a clarification and the editor adds a comment in the code snippet or changes a variable name for clarity.
In this case, such pairs are very specific to the context of the question and cannot be used in recommender systems.
On the other hand, if we find that most of the comments point out errors that the edits fix, then this data is specific to corrective maintenance/bug-fix data sets, as opposed to perfective maintenance for example.
Thus, by understanding the nature of the comments, and accordingly the corresponding edits, we gain a deeper understanding of the potential applications and implications of the extracted pairs. %We also differentiate between useful and not useful pairs.
%This serves two purposes: the first is that it helps us understand if there are generally specific types of comments that result in edit.
%The second is that understanding the differences in the distribution of useful/not useful pairs across these categories may provide helpful information that may allow future work to design heuristics for filtering out not useful pairs.

\renewcommand{\arraystretch}{1.8} 

\begin{table*}[t!]
\centering
\caption{Categories used from Zhang et al.~\protect\cite{ZhangSOComments2019} to label confirmed comment-edit pairs. Note that the listed Stack Overflow IDs are linkable to the answer the comment was addressed to.}
\label{tab:Types}
\resizebox{\textwidth}{!}{
\begin{tabular}{lp{3cm}p{7cm}}
\toprule
Category & Description & Example comment\\
\midrule
%%%%%%%
Correction & 
Provides code correction to the answer & 
\post{10994146}: This gives an undefined variable error. To fix it, change \textasciigrave var\_dump(\$thing);\textasciigrave to \textasciigrave var\_dump(\textbackslash \$thing);\textasciigrave\\
%%%%%%%
Extension  & 
Extends the answer to other cases by making the code more generic, catching corner cases, etc.          
& \post{514517}: One more thing: if you want the range to be inclusive, do \textgreater{}\textgreater{}\textgreater{}for code in range(ord(`a'), ord(`z')+1): print unichr(code) \\
%%%%%%%
Flaw       & 
Points out flaws or limitations. Comments that make small changes but do not change the logic also fall here. e.g., replacing a for loop with a forEach loop & 
\post{2061144}: Don't use query.getSingleResult() as an exception could be thrown if there is not exactly one row returned - see http://java.sun.com/javaee/5/docs/api/javax/ persistence/Query.html\#getSingleResult() \\
%%%%%%%
Error      & 
Points out errors in the code. i.e., incorrect logic resulting in an error or exception & 
\post{39037928}: I tried but it gives error `java.lang.IllegalStateException: You need to use a Theme.AppCompat theme' on setContentView(R.layout.activity home screen); \\
%%%%%%%
Obsolete   &
Points out obsolete APIs, libraries etc. & 
\post{24964658}: While this answer works and seems correct, it was written in 2014 and is now outdated. From Angular 1.4 there is a built in way to do it by using \$httpParamSerializer. Check the answers below for an explanation and an example. \\
%%%%%%%
Disagree   & 
Disagrees with the answer by clarifying the needed requirements. i.e., the answer does not actually answer the question & 
\post{40813524}: But I really need to set the variable at componentDidMount() because it's an object that depends on DOM elements \\ 
%%%%%%%
Question   & 
Asks clarification question about the answer & 
\post{15976303}: So then knownWordsArrayList = new ArrayList\textless{}String\textgreater{}(h); leaves me with all the new words?\\
 %%%%%%%
Request    &
Requests information that is outside the initial question. e.g., follow up questions or asking for an example &
 \post{40611808}: path\_image is a string value.How to set that string value to setBackgroundResourse() \\
 %%%%%%%
Solution   & 
Provides alternative solutions to the answer & 
\post{55069962}: You could even do something like \textasciigrave td:is([data-test=``specific-location"], [data-test=``specific-location1"]) span\textasciigrave to get something a little more compact. \\ 
\bottomrule
\end{tabular}
}
\end{table*}

\paragraph{Methods.} As mentioned in Section~\ref{sec:related}, Zhang et al.~\cite{ZhangSOComments2019} previously categorized the types of comments that exist on Stack Overflow.
Through open-coding, they derived seven high-level comment types (e.g.,  improvement, inquiry, praise) and 17 subtypes (e.g., support, flaw, reference).
Thus, for consistency, we opt for not re-inventing the wheel by performing open coding and developing new categories ourselves; instead, we reuse their fine-grained subtypes to label our data.
Given that their types cover all comments on Stack Overflow, the pairs we extract naturally fall under a subset of these types.
This also means that some of the types they have do not make sense in our context.
For example, a comment praising or supporting the answer will not likely end up causing an edit.
In Table~\ref{tab:Types}, we show the subset of nine subtypes (referred to as category) that are applicable to our context.
For clarity, we also add an example of a real comment from a comment-edit pair that matches this category, as well as any additional assumptions we made about the category in our coding guidelines which may have not have been clear in the original publication.
Given these categories, we perform closed coding where the two authors independently label each confirmed comment-edit pair and then discuss disagreements.

\renewcommand{\arraystretch}{1.2} 

\begin{table*}[t!]
\centering
\caption{Number of total and useful pairs per category}
\label{tab:Type-counts}
\resizebox{\textwidth}{!}{
\begin{tabular}{@{}lrrrrrrrrrrrr@{}}
\toprule
\multirow{2}{*}{\textbf{Category}}&
\multicolumn{2}{c}{\textbf{Java}}&
\multicolumn{2}{c}{\textbf{JavaScript}}&
\multicolumn{2}{c}{\textbf{Android}}&
\multicolumn{2}{c}{\textbf{Python}}&
\multicolumn{2}{c}{\textbf{Php}}&
\multicolumn{2}{c}{\textbf{Overall}}\\
\cmidrule(l){2-3} \cmidrule(l){4-5}\cmidrule(l){6-7}\cmidrule(l){8-9}\cmidrule(l){10-11}\cmidrule(l){12-13}
&\textbf{All}&
\textbf{Useful}&
\textbf{All}&
\textbf{Useful}&
\textbf{All}&
\textbf{Useful}&
\textbf{All}&
\textbf{Useful}&
\textbf{All}&
\textbf{Useful}&
\textbf{All}&
\textbf{Useful}
\\
\midrule
%%%%%%%%%%%%%%%%%%%%%%%%%
\textbf{Error}  & 
%Java
98 & 22 (22\%)
%Javascript
& 91 & 21 (23\%) 
%Android
& 126 & 42 (33\%) 
%Python                                             
& 88 & 35 (40\%)        
%Php                                         
& 108  & 17 (16\%)       
%All                                     
&511 & 137 (27\%)  \\
%%%%%%%%%%%%%%%%%%%%%%%%%
\textbf{Request} & 
%Java
60 & 1 (2\%)
%Javascript
& 53 & 1 (2\%) 
%Android
& 44 & 1 (2\%) 
%Python                                             
& 34 & 0 (0\%)        
%Php                                         
& 45  & 0 (0\%)       
%All                                     
&236& 3 (1\%)  \\
%%%%%%%%%%%%%%%%%%%%%%%%%
\textbf{Correction}& 
%Java
23 & 9 (39\%)
%Javascript
& 47 & 34 (72\%) 
%Android
& 26  & 17 (65\%) 
%Python                                             
& 52 & 43 (83\%)        
%Php                                         
& 51  & 30 (58\%)       
%All                                     
&199 & 133 (67\%)\\
%%%%%%%%%%%%%%%%%%%%%%%%%
\textbf{Disagree} &
 %Java
39 & 2 (5\%)
%Javascript
& 31 & 1 (3\%) 
%Android
& 35 & 0 (0\%) 
%Python                                             
& 42 & 0 (0\%)        
%Php                                         
& 37  & 0 (0\%)       
%All                                     
&184 & 3 (2\%)  \\
%%%%%%%%%%%%%%%%%%%%%%%%%
\textbf{Question} & 
 %Java
35 & 4 (11\%)
%Javascript
& 35 & 5 (14\%) 
%Android
& 28 & 3 (11\%) 
%Python                                             
& 21 & 3 (14\%)        
%Php                                         
& 24  & 1 (4\%)       
%All                                     
&143 & 16 (11\%)  \\
%%%%%%%%%%%%%%%%%%%%%%%%%
\textbf{Flaw} & 
%Java
22 & 12 (55\%)
%Javascript
& 21 & 11 (52\%) 
%Android
& 5  & 3 (60\%) 
%Python                                             
& 20 & 13 (65\%)        
%Php                                         
& 11  & 8 (73\%)       
%All                                     
&79 & 47 (59\%) \\
%%%%%%%%%%%%%%%%%%%%%%%%%
\textbf{Solution} & 
%Java
22 & 13 (59\%)
%Javascript
& 11 & 8 (73\%) 
%Android
& 8 & 2 (25\%) 
%Python                                             
& 22 & 8 (36\%)        
%Php                                         
& 8  & 3 (38\%)       
%All                                     
&71& 34 (48\%)  \\ 
%%%%%%%%%%%%%%%%%%%%%%%%%
\textbf{Extension} & 
%Java
3 & 3 (100\%)
%Javascript
& 13 & 10 (77\%) 
%Android
& 2  & 2 (100\%) 
%Python                                             
& 9 & 2 (22\%)        
%Php                                         
& 2  & 0 (0\%)       
%All                                     
&29 & 17 (59\%) \\
%%%%%%%%%%%%%%%%%%%%%%%%%
\textbf{Obsolete} & 
%Java
1 & 1 (100\%)
%Javascript
& 2 & 0 (0\%) 
%Android
& 2 & 1 (50\%) 
%Python                                             
& 3 & 3 (100\%)        
%Php                                         
& 1  & 1 (100\%)       
%All                                     
&9 & 6 (67\%)  \\
%%%%%%%%%%%%%%%%%%%%%%%%%
\textbf{Other} & 
%Java
2 & 0 (0\%)
%Javascript
& 3 & 0 (0\%) 
%Android
& 8 & 0 (0\%) 
%Python                                             
& 1 & 0 (0\%)        
%Php                                         
& 7  & 0 (0\%)       
%All                                     
& 21 & 0 (0\%)  \\ 
\midrule
\textbf{Total} & 
%Java
305 & 67 (22\%)
%Javascript
& 307 & 91 (30\%) 
%Android
& 284 & 71 (25\%) 
%Python                                             
& 292 & 107 (37\%)        
%Php                                         
& 294  & 60 (20\%)  
%All                                     
&1,482& 396 (27\%)  \\     
\bottomrule
\end{tabular}}

\end{table*}

\paragraph{Results} Our inter-rater agreement for the closed coding task, measured using Cohen's Kappa, ranges from \checkNum{0.82 - 0.95} and is \checkNum{0.88} overall.
Table~\ref{tab:Type-counts} shows the number of comment-edit pairs in each category, per tag.
For now, we focus on the \textit{All} column which shows the categories across all confirmed pairs in each tag (and overall in the last column). 
From the overall numbers (which are also consistent with the individual tag numbers), the most frequent type of comment-edit pairs is  the \textit{Error} category, followed by \textit{Request}, and \textit{Correction}. 
This is good news since the pairs of type \textit{Error} and \textit{Correction} could potentially be used for automated bug-fix recommendations or other applications related to corrective maintenance. 
We further examine the usefulness of these pairs in \ref{rq:usefulness}. 

It is interesting to see that pairs of type \textit{Question} (\checkNum{143} total pairs) are also frequent.
As shown in the example in Table~\ref{tab:Types}, a comment of category \textit{Question} asks clarifications about the already posted solution, such as asking what a specific statement is doing, or why is there a need to call a specified method call.
The edit usually improves the code snippet to answer that question and/or provides additional textual explanation.
This is interesting, because it conveys that users on Stack Overflow want more information regarding the answer in order to have a deeper understanding of how the answer addresses the question.
While these comments are not useful by the definitions we use in this paper, since they are not self explanatory, their relatively high edit response rate suggest that they result in a quality enhancement of the answer and associated code snippet, in order to make the code more self-explanatory or properly documented. 
%The fact that these comments have a relatively high edit response, regardless of its usefulness to the goal of this project, means that users on Stack Overflow also want others to use and understand their answers.
%This shows that Stack Overflow users in general do not only want solutions for questions but also want themselves and others to fully comprehend the answers given.
%\sn{I don't understand the following sentence}
%\hk{I changed the next few sentences to hopefully convey my point better}
%This means that a dataset using Stack Overflow data will evolve in a way that traditional datasets do not.
%Edits made to the answer's code that address a \textit{Question} comment will improve the answer by changing it such that the code itself either has explanatory comments or is more \textit{self-documenting}.
%These types of changes are uncommon in projects as established code is typically unchanged unless there is a bug fix or additional functionality is needed.
%If there is a commit that performs this kind of code quality enhancement it will typically be in other areas of a project as well, leading to a tangled commit.

The number of pairs of type \textit{Extension} and \textit{Obsolete} are low. This is consistent with Zhang et al.~\cite{ZhangSOComments2019} findings where they find that only 0.8\% of the comments they analyze are of type \textit{extension} and 1.0\% are of type \textit{obsolete}.
However, it is interesting to note that these types of pairs are related to perfective maintenance, which opens the door for new types of code recommender systems. 
 
\begin{findingenv}{\ref{rq:changetype}}{rq3}
The most common categories for the extracted comment-edit pairs are \textit{Error}, followed by \textit{Request}, and \textit{Correction}. 
\end{findingenv}
\section{\ref{rq:usefulness}: Usefulness of Comment-Edit Pairs}
\label{sec:rq4}

So far, we have shown that the precision of the extracted pairs is high (i.e., the comment is really related to the edit), the majority of the edits are not tangled, and that the types of comments and changes are promising for various software engineering applications related to code maintenance activities.
However, it is still not clear if these pairs are actually \textit{useful} in the end. 
This is what we investigate in this last research question.

\paragraph{Methods.} As part of our labeling, we also record the usefulness of the \checkNum{1,482} confirmed pairs. As mentioned in the introduction, we consider a pair as \textit{useful} if (1) the edit happens to an existing code snippet in the answer and (2) if the comment describes this change in a way that is understandable outside of the posted Stack Overflow question. 
The first criterion stems from how \maintenance data sets are typically used.
For example, the before version of a bug-fix can be matched to existing code in a repository and the after version is then recommended or automatically applied.
Thus, the first criterion ensures that there is a before version of a code snippet such that it can potentially be compared to existing code.
The second criterion focuses on the comment and ties to our motivation of providing an explanation along with the recommended change.
Instead of just notifying a developer of a potential change to their code, it would be more useful to tell them why this change is needed.
This means that the comment must be understandable on its own without relying on the original thread context.
Again, the two authors independently label the usefulness of the \checkNum{1,482} confirmed pairs and discuss any disagreements.

Finally, to provide external validation for the pairs we mark as useful, we select a total of 15 comment-edit pairs and submit corresponding pull requests.
For the selection of these 15 pairs, our goal was to include pairs from each analyzed SO tag and each pair category.
At the same time, we look for pairs that are simple enough for us to manually implement and create a pull request. 
For example, some pairs identified detailed fixes that would require in depth refactoring and design deliberations by the target repository maintainers.
As such, we selected simple comment-edit pairs, as we want to use these pull requests for additional external validation and confidence, rather than a comprehensive proof of usability.
Table~\ref{tab:pr-descriptive} provides descriptive statistics of these 15 comment-edit pairs.
We wrote a script\footnote{\href{https://github.com/ualberta-smr/QueryGitHub}{https://github.com/ualberta-smr/QueryGitHub}} that uses the GitHub search API to find repositories that match the following criteria:
\begin{enumerate}
\item The repository's main programming language matches that of the tag
\item The repository was active in the last 90 days (i.e., a pushed commit)
\item The repository has at least five stars
\item The repository has at least one closed pull request
\end{enumerate}
These criteria help find active repositories with a higher likelihood of having our pull requests reviewed.
After finding these potential repositories, the script then searches each file in these repositories to find exact code matches of the ``before'' version of the target comment-edit pair.
We manually check any identified files to make sure that we can propose a change that is similar to the edit of the comment-edit pair.
After finding a promising file, we make a pull request that performs a similar change to that in the edit with the description of the pull request being the exact comment, if possible, or a slightly paraphrased version in order to make it more grammatically correct or understandable in a pull request context. 
For example, on Answer\footlink{https://stackoverflow.com/questions/52517618} 52517618, we paraphrased the comment ``\textit{On Java 7 you can also use new String(bytes, StandardCharsets.UTF\_8); which avoids having to catch the UnsupportedEncodingException}" that caused Edit~\footlink{https://stackoverflow.com/revisions/52517618/4} 4 on the answer, to ``\textit{Using new String(bytes, StandardCharsets.UTF\_8) avoids the possibility of throwing an UnsupportedEncodingException.}" on the description of the pull request made to Apache Beam\footlink{https://github.com/apache/beam/pull/11017}.
We show the details of all the pull requests, including their categories, in Table~\ref{tab:pr-details}. Our artifact page~\cite{artifact} also contains the details and links of all our submitted pull requests.

\begin{table}[]
\centering
\caption{Categories and tags of the 15 comment-edit pairs used to make pull requests}
\begin{tabular}{@{}lrrrrrr@{}} 
\toprule
\multirow{2}{*}{Category} & \multicolumn{5}{c}{Tag} & \multirow{2}{*}{Total} \\
\cmidrule(l){2-6}
%\midrule
     & \multicolumn{1}{l}{Java} & \multicolumn{1}{l}{JavaScript} & \multicolumn{1}{l}{Android} & \multicolumn{1}{l}{Python} & \multicolumn{1}{l}{Php} \\
\toprule
Solution   & 2 & 1 & 0 & 0 & 1 &4 \\
Question   & 0 & 1 & 0 & 0 & 0 & 1 \\
Extension  & 0 & 1 & 0 & 0 & 0 & 1 \\
Flaw       & 1 & 1 & 0 & 1 & 2 & 5\\
Correction & 0 & 0 & 2 & 1 & 0 & 3\\
Obsolete   & 0 & 0 & 0 & 1 & 0 & 1\\
\midrule
Total & 3& 4 & 2 & 3 & 3 & 15\\
\bottomrule                 	
\end{tabular}
\label{tab:pr-descriptive}
\end{table}

\definecolor{caribbeangreen}{rgb}{0.0, 0.8, 0.6}
\definecolor{carminered}{rgb}{1.0, 0.0, 0.22}
\newcommand{\acceptedpr}{\rowcolor{caribbeangreen!20}}
\newcommand{\rejectedpr}{\rowcolor{carminered!20}}
\newcommand{\STAB}[1]{\begin{tabular}{@{}c@{}}#1\end{tabular}}

\begin{table}[]
\caption{Details of submitted pull requests (PR). For each PR, we show the answer and comment it is based on, the category this comment-edit pair belongs to, the repo the PR was submitted to, as well as the actual PR link. Green rows indicate accepted/merged PRs, red rows indicate rejected PRs, and non-highlighted rows are PRs with no response. Comments shown in bold are those that required paraphrasing. The PR link, Answer Id, and Repo columns have links to their respective web page.}
\label{tab:pr-details}
\centering
\resizebox{\textwidth}{!}{
\begin{tabular}{@{}lrlp{6cm}l@{}}
\toprule
\textbf{Category} & \textbf{PR link} & \textbf{Answer Id} & \textbf{Comment}& \textbf{Repo (Stars, Forks)} \\ 
\midrule
\multicolumn{5}{c}{\textbf{Java}}\\
\midrule
\acceptedpr
%\multicolumn{1}{c}{ 
%\multirow{3}{*}{\rotatebox[origin=c]{90}{\parbox[c]{3cm}{\centering Java}}}} 
Solution & \edit{https://github.com/apache/beam/pull/11017}{11017} & \answer{52517618} & \textbf{On Java 7 you can also use `new String(bytes, StandardCharsets.UTF\_8);` which avoids having to catch the `UnsupportedEncodingException`} & \edit{https://github.com/apache/beam}{Apache Beam} (4.2k, 2.7k) \\
%%%%%%%
\acceptedpr
Flaw & \edit{https://github.com/vaadin/framework/pull/11941}{11941} & \answer{32749983} & You should (probably, almost) always use a `StringBuilder` to accumulate strings in a loop, to avoid the performance cost of repeatedly constructing strings. & \edit{https://github.com/vaadin/framework}{Vaadin Framework} (1.6k, 733) \\
%%%%%%%
\rejectedpr
Solution & \edit{https://github.com/openhab/openhab1-addons/pull/5945}{5945} & \answer{5553947} & \textbf{Possibly compare `"true".equalsCaseIgnore(person\_array{[}7{]})` is case it could be `null`, of use `Boolean.parseBoolean(person\_array{[}7{]})`} & \edit{https://github.com/openhab/openhab1-addons}{Openhab1-addons} (3.5k, 1.8k) \\
%%%%%%%
\midrule
\multicolumn{5}{c}{\textbf{JavaScript}}\\
\midrule
%\multirow{4}{*}{Javascript}    
Question & \edit{https://github.com/thinkgem/jeesite/pull/500}{500} & \answer{3180655} & The jQuery doc for `jQuery.data()` (http://api.jquery.com/jQuery.data/) says this is a "low-level method" and that you should use `.data()` instead. Do you know what that means and why? & \edit{https://github.com/thinkgem/jeesite}{Jeesite} (7.5k, 5.9k) \\
%%%%%%%
\acceptedpr
Solution & \edit{https://github.com/PrestaShop/PrestaShop/pull/18314}{18314} & \answer{29842091} & \textbf{Why not using preg\_replace directly? (http://php.net/preg\_replace)} & \edit{https://github.com/PrestaShop/PrestaShop}{PrestaShop} (5.1k, 3.8k) \\
%%%%%%%
\acceptedpr
Extension & \edit{https://github.com/ampproject/amphtml/pull/27175}{27175} & \answer{16578216} & \textbf{Don't forget to include support for browsers that use `.contentDocument` instead of `.contentWindow.document`} & \edit{https://github.com/ampproject/amphtml}{AMP} (13.8k, 3.6k) \\
%%%%%%%
\acceptedpr
Flaw & \edit{https://github.com/kairosdb/kairosdb/pull/617}{617} & \answer{41481803} & `object.hasOwnProperty()` is \_almost never needed\_ in current JS code. A `key in object` test suffices just as well & \edit{https://github.com/kairosdb/kairosdb}{KairosDB} (1.6k, 329) \\
%%%%%%%
\midrule
\multicolumn{5}{c}{\textbf{Android}}\\
\midrule
\acceptedpr
%\multirow{2}{*}{Android}   
Correction & \edit{https://github.com/NativeScript/NativeScript/pull/8464}{8464} & \answer{26933338} & If you have the WRITE\_EXTERNAL\_STORAGE permission you don't need READ\_EXTERNAL\_STORAGE, but yes, he does need WRITE\_EXTERNAL\_STORAGE & \edit{https://github.com/NativeScript/NativeScript}{NativeScript} (19k, 1.4k) \\
%%%%%%%
\acceptedpr
Correction & \edit{https://github.com/Tencent/tinker/pull/1354}{1354} & \answer{33366449} & \textbf{`TextUtils.isEmpty()` is better than using a normal `equals()` since it will also perform a `null` check. This will prevent any error in the future and is a good practice.} & \edit{https://github.com/Tencent/tinker}{Tinker} (15.3k, 3.1k) \\
%%%%%%%
\midrule
\multicolumn{5}{c}{\textbf{Python}}\\
\midrule
\rejectedpr
%\multirow{3}{*}{Python}    
Obsolete & \edit{https://github.com/wistbean/learn\_python3\_spider/pull/15}{15} & \answer{12509737} & `\_\_getslice\_\_` is {[}deprecated since 2.0{]}([link]) in favour of `\_\_getitem\_\_` with a `slice()` argument. & \edit{https://github.com/wistbean/learn_python3_spider}{Learn Python3 Spider} (4.6k, 1.4k) \\
%%%%%%%
\acceptedpr
Correction & \edit{https://github.com/nltk/nltk/pull/2515}{2515} & \answer{35560225} & \textbf{It is not necessary to call keys() in the argument to choice. Iterating over a dict will give you the keys. `a = random.choice(A)` is sufficient (and I think nicer-looking).} & \edit{https://github.com/nltk/nltk}{nltk} (9.4k, 2.4k) \\
%%%%%%%
\acceptedpr
Flaw & \edit{https://github.com/ankitects/anki/pull/525}{525} & \answer{40372658} & \textbf{Some suggestions. Load `kernel32` only once as a module global. In `set`, replace `attrib \textasciicircum 4294967295` with `$\sim$attrib`. In `get`, replace `not not (attrs \& what)` with `bool(attrs \& what)`. } 
& \edit{https://github.com/ankitects/anki}{Anki} (7.1k, 1.1k) \\
%%%%%%%
\midrule
\multicolumn{5}{c}{\textbf{Php}}\\
\midrule
%\multirow{3}{*}{Php}              
Flaw & \edit{https://github.com/gongfuxiang/shopxo/pull/40}{40} & \answer{10341595} & \textbf{+1 would do the same. but `\$word{[}0{]}` would make it even more concise..} & \edit{https://github.com/gongfuxiang/shopxo}{ShopXO} (1.3k, 490) \\
%%%%%%%
\rejectedpr
Solution & \edit{https://github.com/forkcms/forkcms/pull/3063}{3063} & \answer{33191679} & \textbf{Side Node: IMHO using `PREG\_SET\_ORDER` (rather than the default `PREG\_PATTERN\_ORDER`) delivers an easier to process result, cause you simple can `foreach'` the result Array and use single dimensional Access (`{[}1{]}, {[}2{]}, {[}3{]}`) to Access the match Groups. Also with named matchgroups having `$match["link"]` is easier to read than `$matches{[}"link"{]}{[}1{]}` etc.} & \edit{https://github.com/forkcms/forkcms}{Fork CMS} (1.1k, 282) \\
%%%%%%%
\acceptedpr
Flaw & \edit{https://github.com/the-benchmarker/web-frameworks/pull/2425}{2425} & \answer{5013708} & you should check for `\$\_SERVER{[}'HTTPS'{]}` to be set before accessing it. & \edit{https://github.com/the-benchmarker/web-frameworks}{Web-frameworks} (4.5k, 399) \\ %\cmidrule(l){2-9} 
\bottomrule
\end{tabular}
}
\end{table}

\paragraph{Results.} Table~\ref{tab:tangled} shows the descriptive statistics of our useful labeling. Our Cohen's kappa ranged from \checkNum{0.62 - 0.81} across the tags, and is \checkNum{0.74} across all pairs.
Out of the \checkNum{1,482} confirmed pairs, we find only \checkNum{396} (27\%) useful ones. We identify two main reasons for this low percentage.
The first is that in many cases, the edit adds a new code snippet.
For example, a comment points out an alternative way of accomplishing the task or an alternative API to use.
Instead of updating the existing snippet, the edit adds an extra code snippet stating that this is another option to use.
In this case, there is no ``before'' version of this code snippet and thus, it will not satisfy our first criterion.
The second common reason was that the comment is too specific to the commenter's context. 
For example, in Answer\footlink{https://stackoverflow.com/questions/4605982} 4605982, this comment caused an edit: \textit{``layout\_height=``fill\_parent" in combination with layout\_below on ListView and layout\_alignParentBottom on LinearLayout is correct and should work."}
However, the comment is too specific to what the original poster is asking for. 
Not every developer will necessarily want to have that same layout.
Thus, we mark that pair as not useful since it does not make sense outside of the question context.
%This means that the edits must be modifications or additions to existing code snippets in the answer, as opposed to text additions or completely new code snippets.
%For example, if the edit added an example that did not previously exist, then that comment-edit pair would be considered \textbf{not} \textit{useful}, but if an edit added details to a code snippet and did not remove anything, that \textbf{would} be considered \textit{useful})

To better understand the characteristics of the useful pairs, we look deeper into the category information in Table~\ref{tab:Type-counts}.
The second column under every tag shows the number and percentage of the confirmed pairs in the corresponding category that are marked as useful.
The results show that while pairs of type \textit{Error} are the most frequent, only \checkNum{27\%} of them are useful. This is mostly due to the error being specific to the context of the post; for example, reporting that the desired behaviour/functionality is not working correctly. 

On the other hand, the \textit{Correction} category shows both a high frequency and a high percentage of usefulness (67\%).
While pairs of type \textit{Solution}, \textit{Obsolete}, \textit{Extension} and \textit{Flaw} were not frequent, their usefulness was high at 48 - 67\%.
Their high usefulness suggest that if these pairs are presented to a developer, it is likely the recommendation will be taken.

Not surprisingly, the usefulness of pairs of type \textit{Request}, \textit{Disagree}, and \textit{Question} is quite low (1 - 11\%).
Given that the nature of these types of pairs is inherently specific to the post context, it is not surprising that they would not be useful in wider applications.
These results suggest that to increase the potential usefulness of comment-edit pairs, we may need to devise additional techniques that can specifically identify comment-edit pairs in the promising categories. 
We discuss this further in Section~\ref{sec:discussion}.

\label{ref:pr-disc}
Table~\ref{tab:pr-details} shows that out of the 15 pull requests made to unique open source repositories on GitHub, 10 requests have been accepted and merged into their respective repository, two requests are still awaiting responses, and three requests were rejected.
Of the 10 requests that were accepted, five of the comments taken from Stack Overflow needed to be paraphrased. 
As the table shows, we were able to merge contributions into popular repositories with thousands of stars and forks, such as Apache Beam\footnote{https://beam.apache.org/} and NLTK\footnote{https://www.nltk.org/}.

The categories of the accepted PRs were diverse including \textit{Flaw}, \textit{Solution}, \textit{Correction}, and \textit{Extension}.
Pairs of type \textit{Solution} and \textit{Extension} tend to fall under the category of preventative maintenance and these pull requests may serve as an indication of how developers view preventative maintenance code improvements.
Of the four pull requests that were of type \textit{Solution}, two of the pull requests were accepted and the other two were rejected.
One of these requests was rejected because a developer replied that the repository was no longer maintained, while the other request was rejected because they thought that the the alternate solution brought no significant difference to the code.
The pull request related to \textit{Extension} was accepted.
The pair categories \textit{Correction} and \textit{Flaw} belong to corrective maintenance and have a total of seven out of eight pull requests accepted.
This indicates that the pairs retrieved from Stack Overflow have the same value as traditional bug-fix data sets in terms of corrective maintenance.
Although our PRs are clearly not a representative sample, they provide some intuition regarding the potential usefulness and applications of our comment-edit pairs.%intuitively this seems reasonable, as some developers will feel the need to perfect their code while others deem code ``good enough" until necessary to change.

We note that the pull request of type \textit{Question} does not have an obvious relationship to maintenance and is possibly information that is unique to Stack Overflow (repository code is not typically updated because of an asked question).
Unfortunately the pull request has not been responded to yet and is neither accepted or rejected.

Finally, as a note in terms of tangledness of the identified \checkNum{396} useful pairs, only \checkNum{39 (10\%)} of these were tangled. This is aligned with the overall low tangledness of edits on Stack Overflow.

\begin{findingenv}{\ref{rq:usefulness}}{rq4}
Out of \checkNum{1,482} confirmed comment-edit pairs across the five tags, \checkNum{396 (27\%)} were potentially useful. The usefulness of comment-edit pairs varies by category and devising automated techniques to find pairs in promising categories may increase the chances of finding useful pairs.
Additionally, to date, \checkNum{10} out of the \checkNum{15} pull requests we submitted to further demonstrate usefulness were accepted.
\end{findingenv}

\section{Discussion}
\label{sec:discussion}

In this paper, we built tooling to identify comment-edit pairs on Stack Overflow.
Our goal was to investigate if these comment-edit pairs could potentially be used as an additional source of data for \maintenance activities.
One advantage of using Stack Overflow comments is that they may provide a concise explanation for the observed change in the edit.
However, the results from \ref{rq:usefulness} show that while we do find useful pairs, the percentage of these pairs is low at 27\%.
We conclude that while Stack Overflow comment-edit pairs look promising, further improvements to our automated extraction techniques are needed to identify a larger number of useful comment-edit pairs for automated applications.
Since our work is the first to investigate this research direction, our tooling and empirical results provide valuable insights for better leveraging Stack Overflow knowledge to build new data sets. 
Moving forward, the goal would be to find more pairs that are useful in automated applications related to \maintenance activities.
In this section, we discuss our findings and the opportunities and challenges for further extending this line of work.

%Code recommender systems
% coments can be used as explanations
% code snippets are short so problems such as figure out which parts of the changes are related to the bug fix are not there
% error reporting, insufficient answers, and improvements can be used
\subsection{Applications}

\paragraph{Software Engineering Applications.} 

Recent work~\cite{wong2019pythonsyntax} already leverages answer edits for creating data sets of code errors and corrections, but focuses only on syntax errors that are found through compiling various versions of a snippet, and thus does not try to associate reasons for the changes.
As our results in ~\ref{rq:changetype} show, there are many categories of changes that occur in the comment-edit pairs we analyzed, ranging from bug fixes to code style and generalizability improvements in the flaw and extensibility categories.

Our results in Table~\ref{tab:Type-counts} show that the \textit{Error} and \textit{Correction} categories are amongst categories with the highest number of pairs.
Both of these categories fall under corrective maintenance.
Automated techniques for bug detection, bug localization, and program repair provide important corrective maintenance support for developers.
Bug-fix data sets are often used to build~\cite{LuiIICST2013} or evaluate~\cite{Dallmeier:2007} these techniques.
Thus, the \textit{Error} and \textit{Correction} comment-edit pairs can be used to add more data to these data sets.

Table~\ref{tab:Type-counts} also shows that there are several pairs in the \textit{Flaw}, \textit{Obsolete}, \textit{Solution}, and \textit{Extension} categories, which fall under corrective or preventative maintenance respectively.
In total, from the 1,482 confirmed pairs, there are 188 ($\sim13\%$) pairs across these four categories. 
Interestingly, despite not being a high absolute number, these four categories were amongst the highest percentage of Useful pairs (59\%, 48\%, 59\%, and 67\% respectively).
This opens the door for automated applications that recommend \textit{improvements} to the code, rather than only bug fixes. 

Regardless of the specific type of application and code maintenance activity, the fact that a Stack Overflow edit in our data set is accompanied by a corresponding comment means that an explanation can be provided to the developer about why a specific code snippet is problematic or why an alternative method of solving something is recommended.
%This is as opposed to commit messages that are either typically short, not always descriptive, and often link to a bug report or associated issue.
For example, in Answer\footlink{https://stackoverflow.com/questions/26933338} 26933338 from Android, the initial provided answer includes a snippet of the manifest file that includes both \texttt{WRITE\_EXTERNAL\_STORAGE} and \texttt{READ\_EXTERNAL\_STORAGE}. The snippet is then edited to remove the latter permission.
If such a removal is suggested to a developer, it will likely not make sense without a concrete reason. The mined comment that is associated with the edit to this answer is \textit{``If you have the WRITE\_EXTERNAL\_STORAGE permission you don't need READ\_EXTERNAL\_STORAGE [..]''}.
When suggesting a fix to this piece of code, providing this comment can help the developer understand why the fix or suggestion is being made.
We used this comment to make one of the accepted pull requests to NativeScript in Table~\ref{tab:pr-details}.
Finally, our results show that the mined comment-edit pairs rarely have multiple unrelated changes (i.e., \textit{tangled changes}).
Thus, our work opens the door for more focused \maintenance data sets, which may potentially work better for generating automated fix scripts~\cite{gao2015fixing}.

\paragraph{Linked Stack Overflow Edit History} Recently, Stack Overflow introduced a new feature that shows a history symbol \raisebox{-\mydepth}{\fbox{\includegraphics[height=\myheight]{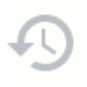}}} beside each question and answer. 
Clicking on this history symbol shows the activity history of the post.
Relating the comments on the post to the edits in the history could be useful to help users understand why an edit was made.
Thus, our matching algorithm can also be applied in that context as future work.

\subsection{Challenges and Opportunities}
\label{sec:challenges-opportunities}

In the above, we discussed the potential applications of using the mined comment-edit pairs.
However, these do not come without challenges since the nature of Stack Overflow data is different than what we traditionally see in version-control systems.
In order to leverage this data source, the ultimate goal is to (automatically) differentiate useful and unuseful pairs.
Such differentiation is difficult for multiple reasons.
We discuss these reasons and potential solutions and/or future work opportunities we perceive.

\paragraph{Non-code comments.}
\label{disc:non-code-comments} 
Our extraction technique favors precision over recall.
Given the amount of answers, edits, and comments on Stack Overflow, we wanted to ensure that we reduce false positives as much as possible.
To do so, we relied on the simple heuristic of focusing on comments that contain code, which allows more precise matching of comments and edits.
This came at the cost of a low recall, as shown in Section~\ref{sec:groundtrutheval}.
Based on our manual investigations on our ground truth data, we find that \textit{non-code comments}, which are comments that contain no code but contain textual descriptions that prompt the answer edit, are one of the main reasons for our low recall (of which an example is also described in Section~\ref{sec:groundtrutheval}).
When considering only comments that contain code, we see that the overall recall of the program rises to 46\%, from  the original 32\%.
One path that could incorporate these non-code comments may be the addition of natural language processing (NLP) techniques that are able to match terminology in the comment and the edit and pair them together. 
For example, one could generate a textual change summary~\cite{GenerateCommit2014} to describe the edit and then match that summary to the comment, while taking into account potential vocabulary mismatch~\cite{ganguly2015word}.
This could potentially enable pairing the explanation in the comment with the changes introduced in the edit even though the comment does not include a code term. 

\paragraph{Conversations.} One challenge we came across during our manual validation is that there is often a conversation occurring in the comments section.
Thus, while many of the comments we have analyzed are stand-alone (recall our second criterion for usefulness), many comments would be difficult to understand without the context of the rest of the conversation.
Such comments would not be useful as explanations provided to users.
The challenge here is to automatically differentiate these two types of comments while extracting comment-edit pairs.
While this is a difficult problem, some ideas from the NLP domain may be potentially useful.
For example, some work looks at automatically inferring context in a sentence~\cite{chan2003dynamic}.
Such techniques can be used to check if the current comment refers to something from the previous comment.
Another simpler technique is to not report comments that were posted within a specific time window (e.g., 30 seconds) from the previous comment.
This is based on our observation that often, a user posts a single big comment split across multiple consecutive ones due to space limitation.

\paragraph{Filler text.} Another challenge related to the mined comments is that some comments are useful and provide a good explanation of the edit, but they contain ``filler'' text.
This includes tagging another participant in the conversation (e.g., a comment from Answer\footlink{https://stackoverflow.com/questions/53216022}
% comment on Nov 8 '18 at 21:06
53216022: \textit{``@Lothar For case-insensitive comparison, use comparing(Contact::getLastName, String.CASE\_INSENSITIVE\_\\ORDER). For language-sensitive comparison, use e.g. comparing(Contact::\\getLastName, Collator.getInstance(Locale.US))"}) or thanking someone for their help (e.g., a comment from Answer\footlink{https://stackoverflow.com/questions/44470955} 44470955,%comment on Jun 10 '17 at 8:08
 \textit{``@binariedM thank but i cant make it work. The console says: ``Uncaught ReferenceError: Invalid left-hand side in assignment" in the line of ``this = x.concat...""}).
In our pull requests, we manually paraphrased comments as needed.
However, ideally, such filler text could be somehow automatically removed. 
Techniques for doing so can be investigated as future work.

\paragraph{Added code.}
\label{disc:added-code} 
Many of the comment-edit pairs we found have helpful suggestions and edits, but unfortunately, the edit is made as an added code snippet. 
This happens especially in the context of the \textit{Solution} category where the answerer typically adds the suggested alternative solution as another code snippet. 
An example of this is found in Answer\footlink{https://stackoverflow.com/questions/20051167} 20051167, which adds the alternative solution provided by the comment: \textit{``If you use substring, then use it till the end: ``0123456789\_".indexOf(check) != -1 No need for matches :)"}.
These pairs are valuable but the main challenge is that there is no ``before'' version, which is why we mark them as not useful. 

Answers may also contain multiple code snippets, for example, to separate steps to be taken or to separate code that should go into multiple files or classes. In these cases, it is not clear which code snippet is being addressed by the added code snippet.
However, added code snippets are typically accompanied by descriptive text, and utilizing these descriptions may provide opportunity to solve this issue (e.g., looking for keywords like ``an alternative is''). 
Accounting for added code may be another opportunity to improve recall of existing comment-edit pairs.
%This may provide an opportunity to improve recall, as opposed to how our current matching technique favors precision over recall and reports only comment-edit pairs where the matched code element occurs in an existing snippet.

\paragraph{Incomplete code.} Many code snippets on Stack Overflow do not include import statements that are necessary to make them compilable or to help in resolving types. Resolving types is necessary for many recommender systems to make use of the comment-edit pairs. This problem has been discussed before in other contexts and there is existing work that tries to infer types for Stack Overflow snippets (e.g.,~\cite{subramanian2014live,SaifullahFQN2019}). That said, one advantage of relying on version-control history, instead of Stack Overflow, is the ability to find tests or containers to reproduce the problem~\cite{tomassi2019bugswarm,dallmeier2007extraction,Just:2014:DDE}.
While specific to Python, there have been recent efforts that attempt to ``dockerize'' a given piece of code found on Stack Overflow or in a GitHub gist~\cite{DockerizeMe2019}.
It would be interesting to see if such efforts can be generalized to allow producing reproducible problems using our extracted comment-edit pairs.  

\paragraph{Pair categories.} We manually categorized our mined pairs. Our results show that some categories have more potential for usefulness than others. Thus, a future opportunity could be automatically categorizing pairs and only reporting pairs that fall in the promising categories. 
Since we share all our data, we foresee future research on designing machine learning classifiers that can automatically assign a category based on specific features of the comment and edit.
While determining these features is not something we explicitly worked on in the context of this work, potential features we foresee from our observations include the size of the edit, the presence of certain keywords (e.g., does not work, error, exception etc), and how many regions/blocks (i.e., text vs. code) have been changed in the edit.
\section{Threats to Validity}

%\todo{mention any threats in the technique, in the generalizability etc}

%This section describes the threats to validity of this project.
As expected with any empirical study, there are several limitations and threats to the validity of our results. We discuss them below.

\paragraph{Construct Validity.}

Since we relied on manual validation to confirm the identified comment-edit pairs, there is a risk that the comments and edits in the pairs we analyze are not actually related.
We mitigate this by defining what a positive label means and by having two authors review the pairs and discuss disagreements.
We also erred on the side of precision and confirmed matches only when we were sure.
We share our exact labeling on our artifact page to facilitate replication and further analysis.

Whether something is useful or not is mostly subjective. In addition to defining an explicit coding guide and having the two authors independently decide on usefulness and discuss disagreements, we also use external validation of usefulness by submitting pull requests to open-source systems based on our data.

\paragraph{Internal Validity}
%Another concern regarding validity would be the regular expressions that were used to find code patterns in both the comments and edits.
The regular expressions we used to identify code terms are taken from Treude et al.~\cite{TreudeRegex}. We modified this list to account for the other languages we analyze and based on experimenting with our ground truth.
However, we cannot claim that the set of regex patterns are complete. While our precision was high, additional regular expressions may potentially catch more comment-edit pairs and improve recall. 

%\sn{what other technical limitations do we have? there must be some but I'm blanking now}
%\hk{The heuristics we used? fuzzy matching?}
% We analyze five tags instead of two tags now. We look at 1910 pairs instead of 399 now. Subjectivity is addressed in "Construct validity". We already talk about the regex patterns. Maybe fuzzy levenshtein is weak? But I don't see any significant difference using a different string matching algorithm.

\paragraph{External Validity}
% Maybe a different title?
A potential threat to the generalizability of our results is that we manually analyze only \checkNum{1,910} pairs.
Even though the sample of \checkNum{1,910} pairs is statistically representative of all detected pairs, the decision to limit the number of pairs to manually analyze was based solely on the amount of labour involved.
The total manual labour involved with the current data is already around \checkNum{129} hours (103 hours for the 1,910 comment-edit pairs and 26 hours to create the ground truth data set), or the equivalent of 16 working days.
Although the authors spent time resolving conflicts and reviewing the analysis, there will always be an element of human bias.

We also analyze only five Stack Overflow tags. While these are popular tags on Stack Overflow and span four different programming languages, our results may not necessarily generalize beyond that. 

Another limitation relates to the pull requests made on open source GitHub repositories.
We make a small number of pull requests (15) which do not establish comprehensive usability of these pairs.
However, the goal of these pull requests was not to be comprehensive but to provide some external validation and confidence in the application of the extracted pairs.  
Although these pull requests provide this confidence, there is inherent bias due to the methods we use to select pairs and find the potential repositories.
Since we used exact code matching in order to find potential repositories instead of a more thorough and precise code parsing approach, we were limited to searching for simple and easily fixable code patterns.
Thus, we do not know how pull requests for more complicated changes might be received by developers.

\section{Conclusion}

%Comments on Stack Overflow provide information to users about any deficiencies of the answer, as well as any improvements that can be made.
%The information provided in comments are important for increasing the quality of Stack Overflow answers and any other crowd-sourced answer forum.
%By knowing the kinds of comments that are causing answers to be updated, multiple systems can be made to raise the quality of answers on the site. 
%Additionally, this data can be used to improve code recommender systems.
%Some of these systems are mentioned in this paper, such as code recommender systems, and possible plugins that could improve answers as a user is writing them.

In this paper, we study comment-edit pairs extracted from Stack Overflow answers.
We implement a technique for identifying comments that resulted in edits to code blocks in the answers.
We run this technique on five popular Stack Overflow tags and share \checkNum{248,399} resulting comment-edit pairs on our artifact page~\cite{artifact}.
We then manually validate a statistically representative sample of \checkNum{1,910} randomly selected comment-edit pairs and confirm \checkNum{1,482} of them.
We then categorize these \checkNum{1,482} pairs and also determine their usefulness and whether the edits are tangled.

We find that the edits are rarely tangled (only 11\%) and that \checkNum{27\%} of the confirmed pairs are useful.
Our results show that categories such as \textit{Correction}, \textit{Extension}, and \textit{Flaw} are particularly useful.
Since we share our data set, future work may explore automatically classifying comment-edit pairs such that only those from promising categories are reported.

We conclude that Stack Overflow is a promising additional source of information for mining \maintenance data sets that can be used in various types of code recommenders and software engineering applications.
However, further work needs to be done to increase the number of extracted useful pairs. 
We presented the current open challenges, such as accounting for non-code comments and added code, as well as some ideas on how future work may address these problems.
We also showed that the type of comments and edits we already find have been useful for getting pull requests merged in popular open-source repositories.
All our data and code are available online~\cite{artifact}.
We hope that this data along with the discussion we provide about future extensions and opportunities encourages further research in this area.

\section*{Acknowledgments}
This research was undertaken thanks to funding from the Canada Research Chair program and from the Natural Sciences and Engineering Research Council. We would also like to thank Sebastian Baltes and Christoph Treude for their feedback regarding the ideas in this work. 

\bibliographystyle{IEEEtran}
\bibliography{references/references}

\begin{thebibliography}{10}
\providecommand{\url}[1]{#1}
\csname url@samestyle\endcsname
\providecommand{\newblock}{\relax}
\providecommand{\bibinfo}[2]{#2}
\providecommand{\BIBentrySTDinterwordspacing}{\spaceskip=0pt\relax}
\providecommand{\BIBentryALTinterwordstretchfactor}{4}
\providecommand{\BIBentryALTinterwordspacing}{\spaceskip=\fontdimen2\font plus
\BIBentryALTinterwordstretchfactor\fontdimen3\font minus
  \fontdimen4\font\relax}
\providecommand{\BIBforeignlanguage}[2]{{%
\expandafter\ifx\csname l@#1\endcsname\relax
\typeout{** WARNING: IEEEtran.bst: No hyphenation pattern has been}%
\typeout{** loaded for the language `#1'. Using the pattern for}%
\typeout{** the default language instead.}%
\else
\language=\csname l@#1\endcsname
\fi
#2}}
\providecommand{\BIBdecl}{\relax}
\BIBdecl

\bibitem{swanson1976dimensions}
E.~B. Swanson, ``The dimensions of maintenance,'' in \emph{Proceedings of the
  2nd international conference on Software engineering}, 1976, pp. 492--497.

\bibitem{fenton1999critique}
N.~E. Fenton and M.~Neil, ``A critique of software defect prediction models,''
  \emph{IEEE Transactions on software engineering}, vol.~25, no.~5, pp.
  675--689, 1999.

\bibitem{amann2018systematic}
S.~Amann, H.~A. Nguyen, S.~Nadi, T.~N. Nguyen, and M.~Mezini, ``A systematic
  evaluation of static {API}-misuse detectors,'' \emph{IEEE Transactions on
  Software Engineering}, 2018.

\bibitem{amann2016mubench}
S.~Amann, S.~Nadi, H.~A. Nguyen, T.~N. Nguyen, and M.~Mezini, ``{MUBench}: {A}
  benchmark for {API}-misuse detectors,'' in \emph{2016 IEEE/ACM 13th Working
  Conference on Mining Software Repositories}.\hskip 1em plus 0.5em minus
  0.4em\relax IEEE, 2016, pp. 464--467.

\bibitem{GazzaloSWRepair}
L.~{Gazzola}, D.~{Micucci}, and L.~{Mariani}, ``Automatic software repair: {A}
  survey,'' \emph{IEEE Transactions on Software Engineering}, vol.~45, no.~1,
  pp. 34--67, 2019.

\bibitem{Just:2014:DDE}
\BIBentryALTinterwordspacing
R.~Just, D.~Jalali, and M.~D. Ernst, ``{Defects4J}: {A} database of existing
  faults to enable controlled testing studies for {Java} programs,'' in
  \emph{Proceedings of the 2014 International Symposium on Software Testing and
  Analysis}, ser. ISSTA '14.\hskip 1em plus 0.5em minus 0.4em\relax New York,
  NY, USA: ACM, 2014, pp. 437--440. [Online]. Available:
  \url{http://doi.acm.org/10.1145/2610384.2628055}
\BIBentrySTDinterwordspacing

\bibitem{dallmeier2007extraction}
V.~Dallmeier and T.~Zimmermann, ``Extraction of bug localization benchmarks
  from history,'' in \emph{Proceedings of the twenty-second IEEE/ACM
  international conference on Automated software engineering}, 2007, pp.
  433--436.

\bibitem{cifuentes2009begbunch}
C.~Cifuentes, C.~Hoermann, N.~Keynes, L.~Li, S.~Long, E.~Mealy, M.~Mounteney,
  and B.~Scholz, ``{BegBunch}: Benchmarking for {C} bug detection tools,'' in
  \emph{Proceedings of the 2nd International Workshop on Defects in Large
  Software Systems: Held in conjunction with the ACM SIGSOFT International
  Symposium on Software Testing and Analysis}, 2009, pp. 16--20.

\bibitem{RaduNFBugs2019}
\BIBentryALTinterwordspacing
A.~Radu and S.~Nadi, ``A dataset of non-functional bugs,'' in \emph{Proceedings
  of the 16th International Conference on Mining Software Repositories}, ser.
  MSR '19.\hskip 1em plus 0.5em minus 0.4em\relax Piscataway, NJ, USA: IEEE
  Press, 2019, pp. 399--403. [Online]. Available:
  \url{https://doi.org/10.1109/MSR.2019.00066}
\BIBentrySTDinterwordspacing

\bibitem{Sliwerski2005}
\BIBentryALTinterwordspacing
J.~\'{S}liwerski, T.~Zimmermann, and A.~Zeller, ``When do changes induce
  fixes?'' in \emph{Proceedings of the 2005 International Workshop on Mining
  Software Repositories}, ser. MSR '05.\hskip 1em plus 0.5em minus 0.4em\relax
  New York, NY, USA: ACM, 2005, pp. 1--5. [Online]. Available:
  \url{http://doi.acm.org/10.1145/1082983.1083147}
\BIBentrySTDinterwordspacing

\bibitem{KimBugIntroChange}
S.~{Kim}, T.~{Zimmermann}, K.~{Pan}, and E.~J. {Jr. Whitehead}, ``Automatic
  identification of bug-introducing changes,'' in \emph{21st IEEE/ACM
  International Conference on Automated Software Engineering}, 2006, pp.
  81--90.

\bibitem{Nguyen:2012}
\BIBentryALTinterwordspacing
A.~T. Nguyen, T.~T. Nguyen, H.~A. Nguyen, and T.~N. Nguyen, ``Multi-layered
  approach for recovering links between bug reports and fixes,'' in
  \emph{Proceedings of the ACM SIGSOFT 20th International Symposium on the
  Foundations of Software Engineering}, ser. FSE '12.\hskip 1em plus 0.5em
  minus 0.4em\relax New York, NY, USA: ACM, 2012, pp. 63:1--63:11. [Online].
  Available: \url{http://doi.acm.org/10.1145/2393596.2393671}
\BIBentrySTDinterwordspacing

\bibitem{Bachmann:2010}
\BIBentryALTinterwordspacing
A.~Bachmann, C.~Bird, F.~Rahman, P.~Devanbu, and A.~Bernstein, ``The missing
  links: {Bugs} and bug-fix commits,'' in \emph{Proceedings of the Eighteenth
  ACM SIGSOFT International Symposium on Foundations of Software Engineering},
  ser. FSE '10.\hskip 1em plus 0.5em minus 0.4em\relax New York, NY, USA: ACM,
  2010, pp. 97--106. [Online]. Available:
  \url{http://doi.acm.org/10.1145/1882291.1882308}
\BIBentrySTDinterwordspacing

\bibitem{Bird:2009}
\BIBentryALTinterwordspacing
C.~Bird, A.~Bachmann, E.~Aune, J.~Duffy, A.~Bernstein, V.~Filkov, and
  P.~Devanbu, ``Fair and balanced?: {Bias} in bug-fix datasets,'' in
  \emph{Proceedings of the 7th Joint Meeting of the European Software
  Engineering Conference and the ACM SIGSOFT Symposium on The Foundations of
  Software Engineering}, ser. ESEC/FSE '09.\hskip 1em plus 0.5em minus
  0.4em\relax New York, NY, USA: ACM, 2009, pp. 121--130. [Online]. Available:
  \url{http://doi.acm.org/10.1145/1595696.1595716}
\BIBentrySTDinterwordspacing

\bibitem{BissyandeBugLink}
T.~F. {Bissyandé}, F.~{Thung}, S.~{Wang}, D.~{Lo}, L.~{Jiang}, and
  L.~{Réveillère}, ``Empirical evaluation of bug linking,'' in \emph{2013
  17th European Conference on Software Maintenance and Reengineering}, 2013,
  pp. 89--98.

\bibitem{Herzig:2013Misclassification}
\BIBentryALTinterwordspacing
K.~Herzig, S.~Just, and A.~Zeller, ``It's not a bug, it's a feature: {How}
  misclassification impacts bug prediction,'' in \emph{Proceedings of the 2013
  International Conference on Software Engineering}, ser. ICSE '13.\hskip 1em
  plus 0.5em minus 0.4em\relax Piscataway, NJ, USA: IEEE Press, 2013, pp.
  392--401. [Online]. Available:
  \url{http://dl.acm.org/citation.cfm?id=2486788.2486840}
\BIBentrySTDinterwordspacing

\bibitem{herzig2013impact}
K.~Herzig and A.~Zeller, ``The impact of tangled code changes,'' in \emph{2013
  10th Working Conference on Mining Software Repositories}.\hskip 1em plus
  0.5em minus 0.4em\relax IEEE, 2013, pp. 121--130.

\bibitem{Herzig2016}
\BIBentryALTinterwordspacing
K.~Herzig, S.~Just, and A.~Zeller, ``The impact of tangled code changes on
  defect prediction models,'' \emph{Empirical Software Engineering}, vol.~21,
  no.~2, pp. 303--336, 2016. [Online]. Available:
  \url{https://doi.org/10.1007/s10664-015-9376-6}
\BIBentrySTDinterwordspacing

\bibitem{maalej2010can}
W.~Maalej and H.-J. Happel, ``Can development work describe itself?'' in
  \emph{2010 7th IEEE working conference on Mining Software
  Repositories}.\hskip 1em plus 0.5em minus 0.4em\relax IEEE, 2010, pp.
  191--200.

\bibitem{dyer2013boa}
R.~Dyer, H.~A. Nguyen, H.~Rajan, and T.~N. Nguyen, ``Boa: {A} language and
  infrastructure for analyzing ultra-large-scale software repositories,'' in
  \emph{Proceedings of the 2013 International Conference on Software
  Engineering}.\hskip 1em plus 0.5em minus 0.4em\relax IEEE Press, 2013, pp.
  422--431.

\bibitem{RastkarSummBugReport2010}
S.~{Rastkar}, G.~C. {Murphy}, and G.~{Murray}, ``Summarizing software
  artifacts: {A} case study of bug reports,'' in \emph{2010 ACM/IEEE 32nd
  International Conference on Software Engineering}, vol.~1, 2010, pp.
  505--514.

\bibitem{BaltesSOTorrent}
\BIBentryALTinterwordspacing
S.~Baltes, C.~Treude, and S.~Diehl, ``{SOTorrent}: {Studying} the origin,
  evolution, and usage of {Stack Overflow} code snippets,'' \emph{CoRR}, vol.
  abs/1809.02814, 2018. [Online]. Available:
  \url{http://arxiv.org/abs/1809.02814}
\BIBentrySTDinterwordspacing

\bibitem{SoniMSR19}
\BIBentryALTinterwordspacing
A.~Soni and S.~Nadi, ``Analyzing comment-induced updates on {Stack Overflow},''
  in \emph{Proceedings of the 16th International Conference on Mining Software
  Repositories}, ser. MSR '19.\hskip 1em plus 0.5em minus 0.4em\relax
  Piscataway, NJ, USA: IEEE Press, 2019, pp. 220--234. [Online]. Available:
  \url{https://doi.org/10.1109/MSR.2019.00044}
\BIBentrySTDinterwordspacing

\bibitem{ZhangSOComments2019}
H.~{Zhang}, S.~{Wang}, T.~{Chen}, and A.~E. {Hassan}, ``Reading answers on
  {Stack Overflow}: not enough!'' \emph{IEEE Transactions on Software
  Engineering}, pp. 1--1, 2019.

\bibitem{artifact}
``Online artifact page,'' \url{https://doi.org/10.5281/zenodo.4458586}.

\bibitem{menzies2012promise}
T.~Menzies, B.~Caglayan, E.~Kocaguneli, J.~Krall, F.~Peters, and B.~Turhan,
  ``The promise repository of empirical software engineering data,'' 2012.

\bibitem{dit2013dataset}
B.~Dit, A.~Holtzhauer, D.~Poshyvanyk, and H.~Kagdi, ``A dataset from change
  history to support evaluation of software maintenance tasks,'' in \emph{2013
  10th Working Conference on Mining Software Repositories}.\hskip 1em plus
  0.5em minus 0.4em\relax IEEE, 2013, pp. 131--134.

\bibitem{ohira2015dataset}
M.~Ohira, Y.~Kashiwa, Y.~Yamatani, H.~Yoshiyuki, Y.~Maeda, N.~Limsettho,
  K.~Fujino, H.~Hata, A.~Ihara, and K.~Matsumoto, ``A dataset of high impact
  bugs: Manually-classified issue reports,'' in \emph{2015 IEEE/ACM 12th
  Working Conference on Mining Software Repositories}.\hskip 1em plus 0.5em
  minus 0.4em\relax IEEE, 2015, pp. 518--521.

\bibitem{FERENC2020110691}
\BIBentryALTinterwordspacing
R.~Ferenc, P.~Gyimesi, G.~Gyimesi, Z.~Tóth, and T.~Gyimóthy, ``An
  automatically created novel bug dataset and its validation in bug
  prediction,'' \emph{Journal of Systems and Software}, vol. 169, p. 110691,
  2020. [Online]. Available:
  \url{http://www.sciencedirect.com/science/article/pii/S0164121220301436}
\BIBentrySTDinterwordspacing

\bibitem{tomassi2019bugswarm}
D.~A. Tomassi, N.~Dmeiri, Y.~Wang, A.~Bhowmick, Y.-C. Liu, P.~T. Devanbu,
  B.~Vasilescu, and C.~Rubio-Gonz{\'a}lez, ``Bugswarm: {Mining} and
  continuously growing a dataset of reproducible failures and fixes,'' in
  \emph{2019 IEEE/ACM 41st International Conference on Software
  Engineering}.\hskip 1em plus 0.5em minus 0.4em\relax IEEE, 2019, pp.
  339--349.

\bibitem{EmpricalStudyOfObsoleteAnswers}
H.~{Zhang}, S.~{Wang}, T.~P. {Chen}, Y.~{Zou}, and A.~E. {Hassan}, ``An
  empirical study of obsolete answers on {Stack Overflow},'' \emph{IEEE
  Transactions on Software Engineering}, pp. 1--1, 2019.

\bibitem{ZhangAdaptingCodeExamples}
\BIBentryALTinterwordspacing
T.~Zhang, D.~Yang, C.~Lopes, and M.~Kirnt, ``Analyzing and supporting
  adaptation of online code examples,'' in \emph{Proceedings of the 41st
  International Conference on Software Engineering}, ser. ICSE '19.\hskip 1em
  plus 0.5em minus 0.4em\relax Piscataway, NJ, USA: IEEE Press, 2019, pp.
  316--327. [Online]. Available: \url{https://doi.org/10.1109/ICSE.2019.00046}
\BIBentrySTDinterwordspacing

\bibitem{Barua2014}
\BIBentryALTinterwordspacing
A.~Barua, S.~W. Thomas, and A.~E. Hassan, ``What are developers talking about?
  {An} analysis of topics and trends in {Stack Overflow},'' \emph{Empirical
  Software Engineering}, vol.~19, no.~3, pp. 619--654, 2014. [Online].
  Available: \url{https://doi.org/10.1007/s10664-012-9231-y}
\BIBentrySTDinterwordspacing

\bibitem{rahman2019automatic}
M.~M. Rahman, C.~K. Roy, and D.~Lo, ``Automatic query reformulation for code
  search using crowdsourced knowledge,'' \emph{Empirical Software Engineering},
  vol.~24, no.~4, pp. 1869--1924, 2019.

\bibitem{TruedeICSE16}
C.~{Treude} and M.~P. {Robillard}, ``Augmenting {API} documentation with
  insights from {Stack Overflow},'' in \emph{2016 IEEE/ACM 38th International
  Conference on Software Engineering}, 2016, pp. 392--403.

\bibitem{Subramanian:2014}
\BIBentryALTinterwordspacing
S.~Subramanian, L.~Inozemtseva, and R.~Holmes, ``Live {API} documentation,'' in
  \emph{Proceedings of the 36th International Conference on Software
  Engineering}, ser. ICSE '14.\hskip 1em plus 0.5em minus 0.4em\relax New York,
  NY, USA: ACM, 2014, pp. 643--652. [Online]. Available:
  \url{http://doi.acm.org/10.1145/2568225.2568313}
\BIBentrySTDinterwordspacing

\bibitem{Ponzanelli:2014}
\BIBentryALTinterwordspacing
L.~Ponzanelli, G.~Bavota, M.~Di~Penta, R.~Oliveto, and M.~Lanza, ``Mining
  {Stack Overflow} to turn the {IDE} into a self-confident programming
  prompter,'' in \emph{Proceedings of the 11th Working Conference on Mining
  Software Repositories}, ser. MSR '14.\hskip 1em plus 0.5em minus 0.4em\relax
  New York, NY, USA: ACM, 2014, pp. 102--111. [Online]. Available:
  \url{http://doi.acm.org/10.1145/2597073.2597077}
\BIBentrySTDinterwordspacing

\bibitem{Lin:2019:PMO}
\BIBentryALTinterwordspacing
B.~Lin, F.~Zampetti, G.~Bavota, M.~Di~Penta, and M.~Lanza, ``Pattern-based
  mining of opinions in {Q\&A} websites,'' in \emph{Proceedings of the 41st
  International Conference on Software Engineering}, ser. ICSE '19.\hskip 1em
  plus 0.5em minus 0.4em\relax Piscataway, NJ, USA: IEEE Press, 2019, pp.
  548--559. [Online]. Available: \url{https://doi.org/10.1109/ICSE.2019.00066}
\BIBentrySTDinterwordspacing

\bibitem{LiuSANER18}
X.~{Liu} and H.~{Zhong}, ``Mining {Stack Overflow} for program repair,'' in
  \emph{2018 IEEE 25th International Conference on Software Analysis, Evolution
  and Reengineering}, 2018, pp. 118--129.

\bibitem{wong2019pythonsyntax}
A.~W. {Wong}, A.~{Salimi}, S.~{Chowdhury}, and A.~{Hindle}, ``Syntax and {Stack
  Overflow}: {A} methodology for extracting a corpus of syntax errors and
  fixes,'' in \emph{2019 IEEE International Conference on Software Maintenance
  and Evolution}, 2019, pp. 318--322.

\bibitem{thiselton2019compiler}
\BIBentryALTinterwordspacing
E.~Thiselton and C.~Treude, ``Enhancing {Python} compiler error messages via
  {Stack Overflow},'' \emph{CoRR}, vol. abs/1906.11456, 2019. [Online].
  Available: \url{http://arxiv.org/abs/1906.11456}
\BIBentrySTDinterwordspacing

\bibitem{gao2015fixing}
Q.~Gao, H.~Zhang, J.~Wang, Y.~Xiong, L.~Zhang, and H.~Mei, ``Fixing recurring
  crash bugs via analyzing {Q\&A} sites (t),'' in \emph{2015 30th IEEE/ACM
  International Conference on Automated Software Engineering}.\hskip 1em plus
  0.5em minus 0.4em\relax IEEE, 2015, pp. 307--318.

\bibitem{DBLP:conf/kbse/FalleriMBMM14}
\BIBentryALTinterwordspacing
J.~Falleri, F.~Morandat, X.~Blanc, M.~Martinez, and M.~Monperrus,
  ``Fine-grained and accurate source code differencing,'' in \emph{{ACM/IEEE}
  International Conference on Automated Software Engineering, {ASE} '14,
  Vasteras, Sweden - September 15 - 19, 2014}, 2014, pp. 313--324. [Online].
  Available: \url{http://doi.acm.org/10.1145/2642937.2642982}
\BIBentrySTDinterwordspacing

\bibitem{AdajiEditComment}
\BIBentryALTinterwordspacing
I.~Adaji and J.~Vassileva, ``Modelling user collaboration in social networks
  using edits and comments,'' in \emph{Proceedings of the 2016 Conference on
  User Modeling Adaptation and Personalization}, ser. UMAP '16.\hskip 1em plus
  0.5em minus 0.4em\relax New York, NY, USA: ACM, 2016, p. 111–114. [Online].
  Available: \url{https://doi.org/10.1145/2930238.2930289}
\BIBentrySTDinterwordspacing

\bibitem{EditBadgeGamification}
S.~Wang, T.-H.~P. Chen, and A.~E. Hassan, ``How do users revise answers on
  technical {Q\&A} websites? {A} case study on {Stack Overflow},'' \emph{IEEE
  Transactions on Software Engineering}, vol.~PP, pp. 1--1, 2018.

\bibitem{Dalip:2013:EUF:2484028.2484072}
\BIBentryALTinterwordspacing
D.~H. Dalip, M.~A. Gon\c{c}alves, M.~Cristo, and P.~Calado, ``Exploiting user
  feedback to learn to rank answers in {Q\&A} forums: A case study with {Stack
  Overflow},'' in \emph{Proceedings of the 36th International ACM SIGIR
  Conference on Research and Development in Information Retrieval}, ser. SIGIR
  '13.\hskip 1em plus 0.5em minus 0.4em\relax New York, NY, USA: ACM, 2013.
  [Online]. Available: \url{http://doi.acm.org/10.1145/2484028.2484072}
\BIBentrySTDinterwordspacing

\bibitem{Diamantopoulos:2019:TMA:3341883.3341922}
\BIBentryALTinterwordspacing
T.~Diamantopoulos, M.-I. Sifaki, and A.~L. Symeonidis, ``Towards mining answer
  edits to extract evolution patterns in {Stack Overflow},'' in
  \emph{Proceedings of the 16th International Conference on Mining Software
  Repositories}, ser. MSR '19.\hskip 1em plus 0.5em minus 0.4em\relax
  Piscataway, NJ, USA: IEEE Press, 2019, pp. 215--219. [Online]. Available:
  \url{https://doi.org/10.1109/MSR.2019.00043}
\BIBentrySTDinterwordspacing

\bibitem{ToxicCode2019}
C.~{Ragkhitwetsagul}, J.~{Krinke}, M.~{Paixao}, G.~{Bianco}, and R.~{Oliveto},
  ``Toxic code snippets on {Stack Overflow},'' \emph{IEEE Transactions on
  Software Engineering}, pp. 1--1, 2019.

\bibitem{RaoDaumClarificationQuestions}
\BIBentryALTinterwordspacing
S.~Rao and H.~D. III, ``Learning to ask good questions: {Ranking} clarification
  questions using neural expected value of perfect information,'' \emph{CoRR},
  vol. abs/1805.04655, 2018. [Online]. Available:
  \url{http://arxiv.org/abs/1805.04655}
\BIBentrySTDinterwordspacing

\bibitem{JinEditsOnHighlyAnsweredQuestions}
\BIBentryALTinterwordspacing
X.~Jin and F.~Servant, ``What edits are done on the highly answered questions
  in {Stack Overflow}?: An empirical study,'' in \emph{Proceedings of the 16th
  International Conference on Mining Software Repositories}, ser. MSR
  '19.\hskip 1em plus 0.5em minus 0.4em\relax Piscataway, NJ, USA: IEEE Press,
  2019, pp. 225--229. [Online]. Available:
  \url{https://doi.org/10.1109/MSR.2019.00045}
\BIBentrySTDinterwordspacing

\bibitem{mchugh2012interrater}
M.~L. McHugh, ``Interrater reliability: {The} kappa statistic,''
  \emph{Biochemia medica: Biochemia medica}, vol.~22, no.~3, pp. 276--282,
  2012.

\bibitem{BaltesEditHistory}
\BIBentryALTinterwordspacing
S.~Baltes, ``Edit and comment history of {Stack Overflow} threads,'' 2018.
  [Online]. Available:
  \url{https://empirical-software.engineering/blog/sotorrent-edithistory}
\BIBentrySTDinterwordspacing

\bibitem{Levenshtein_SPD66}
V.~I. Levenshtein, ``Binary codes capable of correcting deletions, insertions
  and reversals.'' \emph{Soviet Physics Doklady}, vol.~10, no.~8, pp. 707--710,
  1966, doklady Akademii Nauk SSSR, V163 No4 845-848 1965.

\bibitem{fuzzywuzzy}
\BIBentryALTinterwordspacing
ChairNerd, ``{Fuzzywuzzy: Fuzzy string matching in {Python}},'' (2011, July
  08). [Online]. Available:
  \url{https://chairnerd.seatgeek.com/fuzzywuzzy-fuzzy-string-matching-in-python/}
\BIBentrySTDinterwordspacing

\bibitem{viera2005understanding}
A.~J. Viera, J.~M. Garrett \emph{et~al.}, ``Understanding interobserver
  agreement: {The} kappa statistic,'' \emph{Fam med}, vol.~37, no.~5, pp.
  360--363, 2005.

\bibitem{LuiIICST2013}
C.~{Liu}, J.~{Yang}, L.~{Tan}, and M.~{Hafiz}, ``R2fix: {Automatically}
  generating bug fixes from bug reports,'' in \emph{2013 IEEE Sixth
  International Conference on Software Testing, Verification and Validation},
  2013, pp. 282--291.

\bibitem{Dallmeier:2007}
\BIBentryALTinterwordspacing
V.~Dallmeier and T.~Zimmermann, ``Extraction of bug localization benchmarks
  from history,'' in \emph{Proceedings of the Twenty-second IEEE/ACM
  International Conference on Automated Software Engineering}, ser. ASE
  '07.\hskip 1em plus 0.5em minus 0.4em\relax New York, NY, USA: ACM, 2007, pp.
  433--436. [Online]. Available:
  \url{http://doi.acm.org/10.1145/1321631.1321702}
\BIBentrySTDinterwordspacing

\bibitem{GenerateCommit2014}
L.~F. {Cortés-Coy}, M.~{Linares-Vásquez}, J.~{Aponte}, and D.~{Poshyvanyk},
  ``On automatically generating commit messages via summarization of source
  code changes,'' in \emph{14th IEEE International Working Conference on Source
  Code Analysis and Manipulation}, 2014, pp. 275--284.

\bibitem{ganguly2015word}
D.~Ganguly, D.~Roy, M.~Mitra, and G.~J. Jones, ``Word embedding based
  generalized language model for information retrieval,'' in \emph{Proceedings
  of the 38th international ACM SIGIR conference on research and development in
  information retrieval}, 2015, pp. 795--798.

\bibitem{chan2003dynamic}
S.~W.~K. Chan and J.~Franklin, ``Dynamic context generation for natural
  language understanding: {A} multifaceted knowledge approach,'' \emph{IEEE
  Transactions on systems, man, and Cybernetics-Part A: Systems and Humans},
  vol.~33, no.~1, pp. 23--41, 2003.

\bibitem{subramanian2014live}
S.~Subramanian, L.~Inozemtseva, and R.~Holmes, ``Live {API} documentation,'' in
  \emph{Proceedings of the 36th International Conference on Software
  Engineering}, 2014, pp. 643--652.

\bibitem{SaifullahFQN2019}
C.~M.~K. {Saifullah}, M.~{Asaduzzaman}, and C.~K. {Roy}, ``Learning from
  examples to find fully qualified names of {API} elements in code snippets,''
  in \emph{2019 34th IEEE/ACM International Conference on Automated Software
  Engineering}, 2019, pp. 243--254.

\bibitem{DockerizeMe2019}
E.~{Horton} and C.~{Parnin}, ``Dockerizeme: Automatic inference of environment
  dependencies for {Python} code snippets,'' in \emph{41st IEEE/ACM
  International Conference on Software Engineering}, 2019, pp. 328--338.

\end{thebibliography}

\end{document}